\documentclass{aa}  

\usepackage{graphicx}
\usepackage{multirow}
\usepackage{subcaption}
\usepackage{amssymb}
\usepackage{tabularx}
\usepackage{fontawesome}
\graphicspath{ {Figures/} }
\usepackage[usenames,dvipsnames]{xcolor}
\usepackage[labelfont=bf]{caption}
\usepackage{orcidlink}
\usepackage{txfonts}
\usepackage{hyperref}
\hypersetup{
    colorlinks=true,
    linkcolor=MidnightBlue,
    filecolor=MidnightBlue,
    citecolor=MidnightBlue,
    urlcolor=MidnightBlue,
}
\usepackage[capitalize,noabbrev]{cleveref}

\newcommand{\bs}[1]{\boldsymbol{#1}}

\newcommand{\sphangle}{\hat{\bs{n}}}

\newcommand{\sphharmonics}[5]{{}_{#1} Y_{#2 #3}^{#4 #5}}

\newcommand{\underoverset}[3]{\underset{#1}{\overset{#2}{#3}}}

\newcommand{\model}{\mathcal{M}}

\newcommand{\drom}{\mathrm{d}}

\begin{document}

   \title{UNIONS-3500 Weak Lensing: IV. 2D cosmological constraints in harmonic space}

   \author{S. Guerrini\orcidlink{0009-0004-3655-4870}
          \inst{1}
          \thanks{\email{sacha.guerrini@cea.fr}}
          \and
          L. W. K. Goh\orcidlink{0000-0002-0104-8132}
          \inst{3,4}
          \and
          F. Hervas-Peters\orcidlink{0009-0008-1839-2969}
          \inst{2,5}
          \and
          C. Daley\orcidlink{0000-0002-3760-2086}
          \inst{2}
          \and
          M. Kilbinger\orcidlink{0000-0001-9513-7138}
          \inst{2}
          \and
          A. Wittje\orcidlink{0000-0002-8173-3438}
          \inst{6}
          \and
          C. Murray\orcidlink{0000-0002-4668-1273},
          \inst{2}
          \and
          L. Baumont\orcidlink{0000-0002-1518-0150},
          \inst{7,8,9}
          \and
          S. Fabbro\orcidlink{0000-0003-2239-7988}
          \inst{10,11}
          \and
          H. Hildebrandt\orcidlink{0000-0002-9814-3338}
          \inst{6}
          \and
          M. J. Hudson\orcidlink{0000-0002-1437-3786}
          \inst{12,13,14}
          \and
          L. van Waerbeke\orcidlink{0000-0002-2637-8728}
          \inst{15}
          \and
          A. H. Wright\orcidlink{0000-0001-7363-7932}
          \inst{6}
          \and
          T. de Boer\orcidlink{0000-0001-5486-2747}
          \inst{16}
          \and
          J.-C. Cuillandre\orcidlink{0000-0002-3263-8645}
          \inst{2}
          \and
          E. Magnier
          \inst{16}
          \and
          A.~W.~McConnachie\orcidlink{0000-0003-4666-6564}
          \inst{10}
          }

   \institute{%
        Universit\'e Paris Cit\'e, Universit\'e Paris-Saclay, CEA, CNRS, AIM, F-91191, Gif-sur-Yvette, France
    \and
        Université Paris-Saclay, Université Paris Cité, CEA, CNRS, AIM, 91191, Gif-sur-Yvette, France 
    \and
        Institute for Astronomy, University of Edinburgh, Royal Observatory, Blackford Hill, Edinburgh EH9 3HJ, UK
    \and
        Higgs Centre for Theoretical Physics, School of Physics and Astronomy, The University of Edinburgh, Edinburgh EH9 3FD, UK
    \and 
        Department of Astronomy, Steward Observatory, University of Arizona, 933 North Cherry Avenue, Tucson, AZ 85721-0065, USA
    \and
        Ruhr University Bochum, Faculty of Physics and Astronomy, Astronomical Institute (AIRUB), German Centre for Cosmological Lensing, 44780 Bochum, Germany
    \and
    Dipartimento di Fisica - Sezione di Astronomia, Università di Trieste, Via Tiepolo 11, 34131 Trieste, Italy
    \and
    INAF-Osservatorio Astronomico di Trieste, Via G. B. Tiepolo 11, 34143 Trieste, Italy
    \and
    IFPU, Institute for Fundamental Physics of the Universe, via Beirut 2, 34151 Trieste, Italy
    \and
        NRC Herzberg Astronomy and Astrophysics, 5071 West Saanich Road, Victoria, BC V8Z 6M7, Canada
    \and
        Department of Computer Science, University of British Columbia, 2366 Main Mall, Vancouver, BC V6T 1Z4, Canada
    \and
        Department of Physics and Astronomy, University of Waterloo, 200 University Avenue West, Waterloo, Ontario N2L 3G1, Canada
    \and
        Waterloo Centre for Astrophysics, University of Waterloo, Waterloo, Ontario N2L 3G1, Canada
    \and
        Perimeter Institute for Theoretical Physics, 31 Caroline St. North, Waterloo, ON N2L 2Y5, Canada
    \and
        Department of Physics and Astronomy, University of British Columbia, 6224 Agricultural Road, V6T 1Z1, Vancouver, Canada
    \and 
         Institute for Astronomy, University of Hawaii, 2680 Woodlawn Drive, Honolulu HI 96822
    }

   \date{Received XXXX; accepted YYYY}
 
  \abstract{
   The Ultraviolet Near Infrared Optical Northern Survey (UNIONS) is a photometric survey in the northern sky. The quality of the data in the $r$ band provides precise shape measurements to measure the growth of structures using cosmic shear.
   This work aims to constrain cosmological parameters using a harmonic-space estimator of the cosmic shear signal, known as pseudo-$C_\ell$, in a non-tomographic analysis.
   We perform our analysis in the context of the standard $\Lambda$CDM cosmology. We model astrophysical systematic effects such as baryonic feedback and intrinsic alignments of galaxies. We verify that the point spread function systematic contribution does not affect our results. We assess the impact of different scale cuts and modelling choices on the constraints.
    We find $S_8 \equiv \sigma_8 \sqrt{\Omega_{\rm m}/0.3} = 0.891^{+0.057}_{-0.084}$, consistent at the $0.79 \, \sigma$ level with \emph{Planck} and between $0.87$ to $1.51 \, \sigma$ with other weak lensing surveys. Our results are robust to analysis choices and we use lognormal simulations to assess the consistency between configuration and harmonic space results, finding a $2.18 \, \sigma$ agreement between the two statistics. The degeneracy between $S_8$ and the amplitude of the intrinsic alignment, $A_{\rm IA}$, sampled from a prior obtained from direct measurements, is one of the largest sources of uncertainty.
    This work is part of the first cosmological analysis of the UNIONS survey using cosmic shear and paves the way for future tomographic and $3 \times 2$ point cross-correlation analyses, exploiting the unique overlap of UNIONS with deep spectroscopic surveys in the northern hemisphere.
   }


   \maketitle
%

\section{Introduction}
Gravitational lensing by the large-scale structure of the Universe distorts the observed shapes of galaxies coherently. This \emph{cosmic shear} effect is sensitive to the growth of structures and the expansion history of the Universe. Cosmic shear is therefore a powerful probe for studying the behaviour of dark energy across time and for testing general relativity on cosmological scales \citep[see, e.g.,][for reviews]{kilbingerCosmologyCosmicShear2015,mandelbaumWeakLensingPrecision2018}. In the last decade, Stage-III photometric surveys such as the Dark Energy Survey \citep[DES;][]{gattiDarkEnergySurvey2021}, the Kilo-Degree Survey \citep[KiDS;][]{wrightFifthDataRelease2024}, and the Hyper-Suprime Cam \citep[HSC;][]{liThreeyearShearCatalog2022} survey have provided strong constraints on the cosmological parameters related to the growth of structure, namely the density of matter $\Omega_{\rm m}$ and the amplitude of structure $\sigma_8$. The constraining power on the combination $S_8 \equiv \sigma_8 \sqrt{\Omega_{\rm m}/0.3}$ is competitive with cosmic microwave background observations \citep{planckcollaborationPlanck2018Results2020}. The use of two-point correlation functions from cosmic shear has become a standard technique to constrain cosmological parameters \citep{amonDarkEnergySurvey2022,seccoDarkEnergySurvey2022, douxDarkEnergySurvey2022, asgariKiDS1000CosmologyCosmic2021, wrightKiDSLegacyCosmologicalConstraints2025, dalalHyperSuprimeCamYear2023,liHyperSuprimeCamYear2023}. Stage-III surveys have paved the way for the next-generation photometric surveys, such as the ESA satellite \textit{Euclid} \citep{mellierEuclidOverviewEuclid2025}, the Vera Rubin Observatory Legacy Survey of Space and Time \citep[LSST;][]{ivezicLSSTScienceDrivers2019} and NASA's Nancy Grace Roman Space Telescope \citep{akesonWideFieldInfrared2019}, which will observe an order of magnitude more galaxies due to improved observations in quality, area, depth, and spectral coverage. Because of their reduced statistical uncertainties, these experiments are pioneering in the study of dark energy and, in particular, of its evolution with time, as suggested by the Dark Energy Spectroscopic Instrument (DESI) first and second Data Releases \citep{desicollaborationDESI2024VI2025,desicollaborationDESIDR2Results2025}. However, the reduced noise level requires an exquisite understanding of astrophysical and instrumental systematics that pollute the cosmic shear signal. This pushes the community to carefully scrutinise each component of the analysis framework, from data reduction to the inference of cosmological parameters. Stage-III surveys have been essential for developing and validating the pipelines that will analyse those next-generation surveys \citep{jeffersonReanalysisStageIIICosmic2025}.

The Ultraviolet Near Infrared Optical Northern Survey \citep[UNIONS;][]{gwynUNIONSUltravioletNearinfrared2025a} is the last of the Stage-III surveys and provides photometric coverage of the northern hemisphere. Its cosmic shear data is complementary to other ground-based surveys, with excellent data quality in the $ugriz$ photometric bands. We can use two-point statistics of the cosmic shear field to extract the Gaussian cosmological information. These are accompanied by a good understanding of their theoretical modelling and different sources of systematic effects \citep{schneiderBaryonicEffectsWeak2020,navarro-gironesPAUSurveyMeasuring2025}. Two-point statistics can also be efficiently measured on the data. The shear two-point function can be expressed in configuration space using the correlation function $\xi_\pm(\vartheta)$ \citep{schneiderAnalysisTwopointStatistics2002}, as a function of angular separation $\vartheta$, or in harmonic space as the shear angular power spectrum, $C_\ell$ \citep{deshpandeEuclidPreparationXXXVI2024}, as a function of multipole $\ell$. While, in principle, the two statistics summarise the same cosmological information \citep{kilbingerPrecisionCalculationsCosmic2017,parkMatchingCosmicShear2025}, their different responses to systematic effects and the nontrivial distribution of cosmological information across angular scales $\vartheta$ and multipoles $\ell$ can lead to differences in the cosmological constraints obtained with one or the other.

In this work, we present measurements of the non-tomographic cosmic shear power spectrum from UNIONS data, which we used to constrain the cosmological parameters of the $\Lambda$CDM model. We assessed the robustness of the modelling by varying scale cuts and modelling choices, such as our choice of the non-linear model for the matter power spectrum and the impact of baryonic feedback, one of the main sources of uncertainty in weak lensing studies. Additionally, we studied the consistency of these constraints with other weak lensing surveys and external probes. We also compared our results to those obtained in configuration space and published in a companion paper \citep{gohUNIONSWeakLensing2026}.

The paper is organised as follows. Section~\ref{sec:UNIONS} presents UNIONS weak lensing data; Section~\ref{sec:methods} introduces the methodology used to measure the non-tomographic cosmic shear power spectrum and to obtain theoretical predictions for given cosmological parameters. The modelling addresses the choices of non-linear matter power spectrum, including baryonic feedback, the intrinsic alignment of galaxies, and the estimation of the covariance of the cosmic shear power spectrum. Section~\ref{sec:inference} presents the inference pipeline and the choice of priors for intrinsic alignment, shear multiplicative bias, and redshift distribution uncertainties. We also discuss the amplitude of the point spread function (PSF) systematics and validate the covariance modelling. Section~\ref{sec:results} presents our main results: cosmological constraints on the growth of structure. We perform robustness checks on our modelling and compare our constraints to those of previous weak lensing experiments and external probes, such as \textit{Planck}. This work is part of a series of papers, which are summarised in Table~\ref{tab:unions_papers}. We report on catalogue construction (Paper~I; \citealt{kilbingerUNIONSWeakLensing2026}); $B$-mode validation (Paper~II; \citealt{daleyUNIONSWeakLensing2026}); configuration space cosmological constraints (Paper~III; \citealt{gohUNIONSWeakLensing2026}); harmonic space cosmological constraints (Paper~IV; This work); and image simulations and validation (Paper~V; \citealt{hervaspetersUNIONSWeakLensing2026}).

\begin{table*}
\centering
\begin{tabular}{l l l}
\hline
\hline
 & \textbf{Author} & \textbf{Title} \\
\hline
I & \cite{kilbingerUNIONSWeakLensing2026} & A Galaxy Shape Catalogue in the Northern Sky \\
II & \cite{daleyUNIONSWeakLensing2026} & $B$-mode validation and comparison for cosmic shear\\
III & \cite{gohUNIONSWeakLensing2026} & 2D cosmological constraints in configuration space\\
IV & \textbf{This work} & 2D cosmological constraints in harmonic space \\
V & \cite{hervaspetersUNIONSWeakLensing2026} & Shear calibration with realistic image simulations \\
\hline
\end{tabular}
\caption{List of associated publications in this coordinated UNIONS release.}
\label{tab:unions_papers}
\end{table*}

\section{The UNIONS data set}
\label{sec:UNIONS}

UNIONS is a photometric survey in the northern sky \citep{gwynUNIONSUltravioletNearinfrared2025a}. It combines multi-band photometric images from telescopes located in Hawai'i to observe $6250$ deg$^2$ of sky upon completion. The Canada-France-Hawai'i Telescope (CFHT) provides $u$- and $r$-band images; this part of the survey is called the Canada-France Imaging Survey (CFIS). The Panoramic Survey Telescope and Rapid Response System (Pan-STARRS) provides $i$- and $z$-band data. Subaru takes images in the $z$-band in the framework of WISHES (Wide Imaging with Subaru HSC of the Euclid Sky), and the $g$-band with WHIGS (Waterloo-Hawai'i IfA $g$-band Survey). UNIONS is a key piece of the \textit{Euclid} survey. The broad filter band of the VIS instrument aboard \textit{Euclid} prevents accurate photometric redshift estimation \citep{euclidcollaborationEuclidIIVIS2025}. The need for ground-based data within the \textit{Euclid} survey footprint necessitated the surveys introduced above and accelerated their progression \citep{scaramellaEuclidPreparationEuclid2022a}. In that context, UNIONS has been granted an extension to collect data down to 15° in declination to close the gap to the planned LSST footprint. However, because the multi-band imaging is incomplete, the analysis in this paper relies on non-tomographic cosmic shear measurements over $3500$ deg$^2$ using CFIS $r$-band data.

\subsection{UNIONS--$3500$ deg$^2$ catalogue}
\label{sec:UNIONS_catalogue}
We use the fiducial UNIONS weak lensing\footnote{corresponding to the version SP v1.4.6.3 in \cite{kilbingerUNIONSWeakLensing2026}.} sample of galaxies introduced in the catalogue paper \citep{kilbingerUNIONSWeakLensing2026}. It is composed of $61\,378\,891$ galaxies covering $A_\mathrm{eff} = 2894$ deg$^2$ of effective area on the sky after masking. The effective density of galaxies amounts to $n_\mathrm{eff} = 4.96$ per square arcminute. The PSF fitting and the shape measurement are performed with \texttt{ShapePipe} \citep{guinotShapePipeNewShape2022, farrensShapePipeModularWeaklensing2022}, and the PSF is modelled using \texttt{PSFex} \citep{bertinAutomatedMorphometrySExtractor2011}. The shape of the galaxies is measured using \texttt{ngmix} \citep{sheldonNGMIXGaussianMixture2015} and the calibration of the galaxy ellipticities is performed using \texttt{Metacalibration} \citep{huffMetacalibrationDirectSelfCalibration2017, sheldonPracticalWeaklensingShear2017} to estimate the shear response matrix, $R$, of each object. The PSF estimation and shape measurement are performed using photometric data from the $r$ band, benefiting from a competitive seeing of approximately $0.7\arcsec$. The weak lensing sample is selected after applying masks to remove areas of the sky polluted by stellar spikes, satellite trails, or other instrumental spurious effects. In addition, we remove small objects, more sensitive to leakage and other PSF effects, applying a cut in size using the ratio of the half-light radius of the galaxy, $r_\mathrm{HLR, gal}$, and of the PSF, $r_\mathrm{HLR, PSF}$: $r_\mathrm{HLR, gal}/r_\mathrm{HLR, PSF} > 0.707$. The exhaustive list of cuts applied is documented in \cite{kilbingerUNIONSWeakLensing2026}.  The output catalogue contains two sets of ellipticities for each galaxy. Calibrated shapes using \texttt{Metacalibration} are saved, but undergo a second correction step to remove residual PSF leakage empirically, adapted from \cite{liKiDSLegacyCalibrationUnifying2023}. More details on the PSF fitting and validation, shape measurement, sample selection, and leakage correction can be found in the associated catalogue paper \citep{kilbingerUNIONSWeakLensing2026}. The impact of this empirical leakage calibration will be discussed in Sects.~\ref{sec:PSF} and \ref{sec:robustness_modelling_psf}. The shape noise per ellipticity component of the weak lensing sample used in this work amounts to $\sigma_e = 0.27$.

\subsection{Redshift distribution}
\label{sec:redshift distribution}
We estimate the redshift distribution for the UNIONS $r$-band weak lensing source sample using the colour--redshift relation \citep[see Sect.\,3.3 of][for full details]{gohUNIONSWeakLensing2026}. The multi-band photometry is taken from the Canada--France--Hawai'i Telescope Lensing Survey \citep[CFHTLenS;][]{heymansCFHTLenSCanadaFranceHawaiiTelescope2012,erbenCFHTLenSCanadaFranceHawaiiTelescope2013}, which provides deep $ugriz$ data \citep{hildebrandtCFHTLenSImprovingQuality2012}. For $44.2~\mathrm{deg}^2$ in the CFHT W3 field, virtually all CFIS $r$-band detections have CFHTLenS counterparts, and we thus cross-match sources and adopt their associated $ugriz$ magnitudes. This matched sample is used as a representative subsample for the colour--redshift distribution of the entire UNIONS population.

To calibrate the redshift distribution, we compile a spectroscopic sample that was observed with the same CFHTLenS $ugriz$ filters as the matched UNIONS--CFHTLenS sources, including galaxies from DEEP2 \citep{newmanDEEP2GalaxyRedshift2013}, the VIMOS VLT Deep Survey \citep[VVDS;][]{lefevreVIMOSVLTDeep2005}, and the VIMOS Public Redshift Survey \citep[VIPERS;][]{scodeggioVIMOSPublicExtragalactic2018}.
Using the multi-band photometry of the spectroscopic sample, we train a self-organising map (SOM; \citealt{kohonenSelforganizedFormationTopologically1982}) based on the colours and magnitudes of the galaxies \citep{mastersMappingGalaxyColorRedshift2015,wrightPhotometricRedshiftCalibration2020}.

\begin{figure}[h!]
    \centering
    \includegraphics[width=\linewidth]{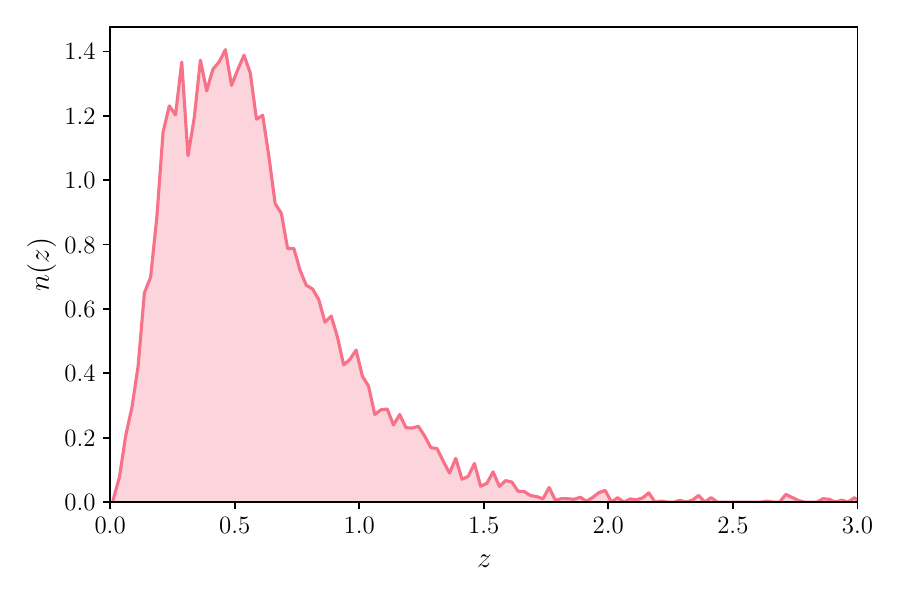}
    \caption{Normalised redshift distribution, $n(z)$. The redshift distribution is obtained using the colour--redshift relation method described in Sect.~\ref{sec:redshift distribution}.}
    \label{fig:nz}
\end{figure}

We populate the SOM with UNIONS sources by assigning each galaxy to its best-matching cell in colour-magnitude. Weights for the SOM, $w_i^{\rm SOM}$, are derived to reweight the spectroscopic calibration sample to match the source sample distribution. They also take into account potential selection biases in both the source and calibration samples using a procedure detailed in Sect.\,3.3 of \cite{gohUNIONSWeakLensing2026}. The redshift distribution of the UNIONS weak-lensing sample is computed by summing over SOM cells:
\begin{equation}
    n(z)
    = \sum_i w_i^{\mathrm{SOM}}\, n_i^{\mathrm{spec}}(z),
\end{equation}
where $n_i^{\mathrm{spec}}(z)$ is the spectroscopic redshift distribution in cell $i$. Figure~\ref{fig:nz} shows the final calibrated distribution $n(z)$.

\section{Methods}
\label{sec:methods}

This work aims to extract cosmological constraints from the measurements of the angular power spectrum of the non-tomographic cosmic shear field inferred from the UNIONS data. This section describes the estimation of the angular power spectrum from the data, the theoretical modelling of the power spectrum, and the covariance matrix used in the multivariate Gaussian likelihood (Sect.~\ref{sec:cov_modelling}).
\subsection{Angular power spectrum measurements}
\label{sec:pseudo_cl}

Cosmic shear is represented by a spin-2 field, $\bs{\gamma} \equiv (\gamma_1, \gamma_2)$. At linear order, it corresponds to the distortions of the observed ellipticities of background galaxies, $\bs{e}^\mathrm{obs} \equiv (e_1^\mathrm{obs}, e_2^\mathrm{obs})$. The galaxies present in the UNIONS galaxy catalogue sample the shear field at discrete positions $\hat{\bs{n}}$ on the celestial sphere.

To constrain cosmology, we estimate the angular power spectrum of the shear field. The true shear field $\bs{\gamma}$ can be decomposed on the basis of spherical harmonics. A spin-2 field can be decomposed into $E$- (curl-free) and $B$- (divergence-free) modes. With $a \in \{1, 2\}$ the component index and $\alpha \in \{E, B\}$ the mode index, this decomposition writes as
\begin{align}
    \gamma^a(\sphangle) = \underset{\ell m}{\sum} \sphharmonics{\pm 2}{\ell}{m}{a}{\alpha}(\sphangle) \gamma_{\ell m}^\alpha,
\end{align}
where $\sphharmonics{s}{\ell}{m}{a}{\alpha}(\sphangle)$ are the generalised spin-$s$ spherical harmonic functions \citep[see, e.g.,][]{hikageShearPowerSpectrum2011} and the component index is a superscript. For full-sky observations, the shear power spectra are then defined by the covariance matrix of the spherical harmonic coefficients,
\begin{align}
 \langle \gamma^\alpha_{\ell m} \gamma^{\beta \, *}_{\ell' m'} \rangle = C_\ell^{\alpha \beta} \delta_{\ell \ell'} \delta_{m m'},
\end{align}
where $\alpha,\beta \in \{E, B\}$, and which can be estimated with
\begin{align}
\label{eq:full-sky estimator}
    \hat{C}_\ell^{\alpha \beta} = \frac 1 {2 \ell +1} \underoverset{m=-\ell}{\ell}{\sum} \gamma^\alpha_{\ell m} \gamma^{\beta \, *}_{\ell m},
\end{align}
where the hat refers to the \emph{full-sky} angular power spectrum throughout. Gravitational lensing does not create $B$ modes to first order. However, a number of effects may generate small $B$-mode power spectra, such as second-order lensing effects \citep{krauseWeakLensingPower2010} or the clustering of source galaxies \citep{schneiderBmodesCosmicShear2002}. Given the statistical sensitivity of UNIONS, these higher-order effects will not create measurable $B$-modes; therefore, $B$-modes are a useful diagnostic to detect potential systematic effects in the data, such as PSF dependence discussed in Sect.~\ref{sec:PSF}. \cite{daleyUNIONSWeakLensing2026} presents a detailed analysis of the $B$-modes estimators used in this study.

This approach works for a full-sky observation, but we have only partial coverage due to the survey footprint. A mask is applied to the shear field that couples multipoles and biases the estimator defined in Eq.~\eqref{eq:full-sky estimator}. We thus estimate the power spectra with the pseudo-$C_\ell$ formalism \citep{hivonMASTERCosmicMicrowave2002} using the \texttt{NaMaster} software \citep{alonsoUnifiedPseudoClFramework2019}. Previous measurements in DES \citep{douxDarkEnergySurvey2022} and HSC \citep{dalalHyperSuprimeCamYear2023} binned the galaxy shape catalogue in pixels on the sphere using \texttt{HealPy} \citep{gorskiHEALPixFrameworkHighResolution2005, zoncaHealpyEqualArea2019} at a chosen resolution and measured the power spectrum of the maps \citep{alonsoUnifiedPseudoClFramework2019}. This approach raises issues at the scale of the pixel resolution or below, where the pixel window function and aliasing can bias the estimation of the pseudo-$C_\ell$. If the source density in the mask is low, it is dominated by Poisson noise, which can lead to numerical instabilities and bias. We address these issues using a catalogue-based estimator \citep{wolzCatalogbasedPseudoCls2025}.

Let $w(\sphangle)$ denote the mask and $\tilde{\bs{\gamma}}(\sphangle) \equiv w(\sphangle) \bs{\gamma}(\sphangle)$ the masked shear field. We will use the subscript $i$ to denote the $i$th point with sky position $\sphangle_i$. The mask and masked shear field are then, using $\delta^{\rm D}(\sphangle_1, \sphangle_2)$ the Dirac delta function on the sphere,
\begin{align}
    w(\sphangle) = \underset{i}{\sum}w_i \delta^{\rm D}(\sphangle, \sphangle_i), \quad \tilde{\bs{\gamma}}(\sphangle) = \underset{i}{\sum} w_i \bs{\gamma}_i \delta^{\rm D}(\sphangle, \sphangle_i).
\end{align}
The power spectrum of the masked shear field has an expectation value of
\begin{align}
\label{eq:pseudo_cl}
    \langle \tilde{C}_\ell^{\alpha \beta} \rangle = \underset{\ell'}{\sum} ({\sf M}_{\ell \ell'})^{\alpha \beta}_{\alpha' \beta'} S_{\ell'}^{\alpha' \beta'} + \tilde{N}_\ell^{\alpha \beta},
\end{align}
where ${\sf M}_{\ell \ell'}$ is the mode-coupling matrix, $S_{\ell}^{\alpha \beta}$ the signal power spectrum of the unmasked shear field and $\tilde{N}_\ell^{\alpha \beta}$ is the noise bias due to shape noise. The tilde notation will be used throughout to refer to quantities obtained from the mask field. The mode-coupling matrix describes how the mask correlates different multipoles as well as the leakage between $E$- and $B$-modes. In addition, one can define an estimator for the binned power spectra, independent of whether the field is masked or not, which is
\begin{align}
    C_L^{\alpha \beta} = \underset{\ell \in L}{\sum} \omega_L^\ell C_\ell^{\alpha \beta},
\end{align}
where $\omega_L^\ell$ is a set of weights defined for multipoles $\ell$ in bandpower $L=[\ell_\mathrm{min}, \ell_\mathrm{max}]$ and normalised such that $\sum_{\ell \in L}\omega_L^\ell = 1$. The estimator for the binned power spectrum is then given by
\begin{align}
\label{eq:estimate_power_spectrum}
    \hat{C}_L^{\alpha \beta} = \underset{L'}{\sum}  ([{\sf M}^S_{L L'}]^{-1})^{\alpha \beta}_{\alpha' \beta'} \langle \tilde{S}_{L'}^{\alpha' \beta'} \rangle,
\end{align}
where $\tilde{S}_L^{\alpha \beta}$ is the signal of the masked power spectrum after removing the noise bias, and the mode-coupling matrix, ${\sf M}^S_{L L'}$, is estimated for this noise-subtracted estimator and binned following 
\begin{align}
    {\sf M}^S_{L L'} = \underset{\ell \in L}{\sum} \underset{\ell' \in L'}{\sum} \omega_L^\ell {\sf M}^S_{\ell \ell'}.
\end{align}
These operations are performed by \texttt{NaMaster}, and more details on the algorithm can be found in \cite{lizancosHarmonicAnalysisDiscrete2024} and \cite{wolzCatalogbasedPseudoCls2025}.

\begin{figure}
    \centering
    \includegraphics[width=1\linewidth]{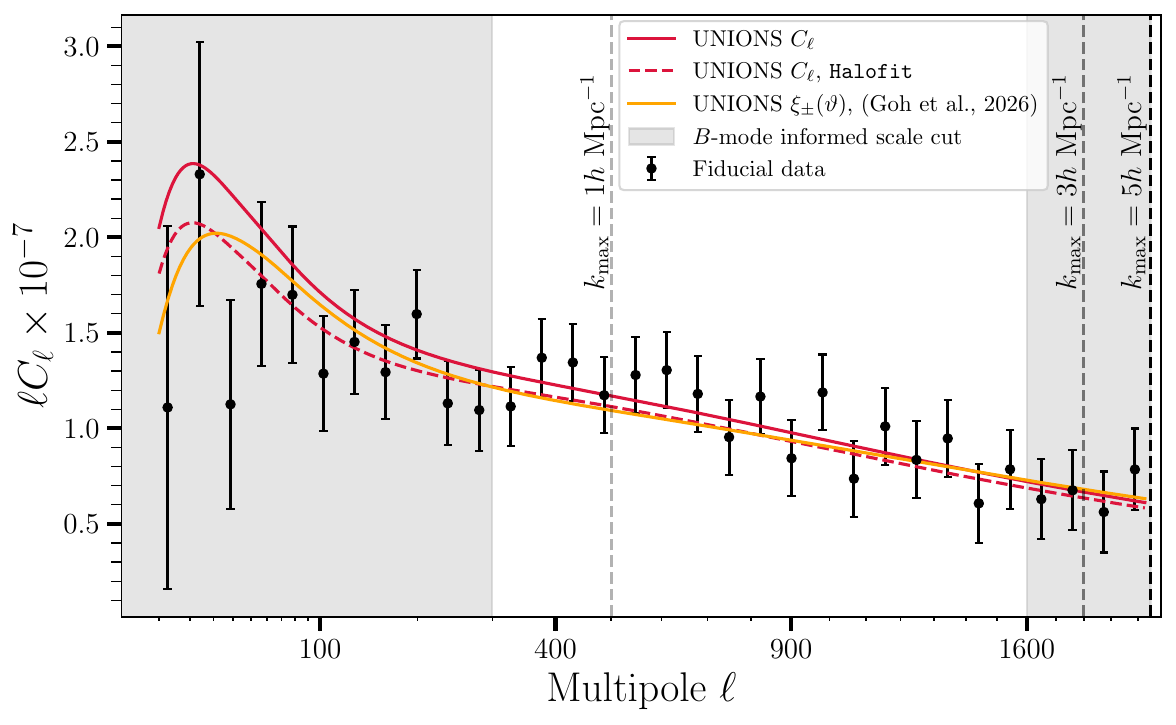}
    \caption{The cosmic shear power spectrum from the UNIONS weak lensing sample. Data points are computed with \texttt{NaMaster} using a catalogue-based estimator (see Sect.~\ref{sec:pseudo_cl}). Error bars are obtained using the Gaussian covariance estimator of \texttt{NaMaster} with added non-Gaussian contributions from \texttt{OneCovariance} (see Sect.~\ref{sec:cov_modelling}). The solid red line corresponds to the best fit obtained using our fiducial setup described in Sect.~\ref{sec:inference}, relying on \texttt{HMCode2020}. The dashed red line is obtained using \texttt{Halofit} for the non-linear power spectrum. The solid orange line corresponds to the best fit obtained in configuration space \citep[see][]{gohUNIONSWeakLensing2026}. The vertical dashed lines represent the scale-cuts applied for different $k_\mathrm{max}$ (see Sect.~\ref{sec:scale_cut}).}
    \label{fig:data_vector_and_best_fit}
\end{figure}

In this work, we apply the same binning strategy as \cite{douxDarkEnergySurvey2022}, using an equal-weight binning scheme with 32 square-root-spaced bins defined between multipoles $\ell_{\rm min}=8$ and $\ell_\mathrm{max} = 2048$. Contrary to \cite{douxDarkEnergySurvey2022}, we do not need to remove the noise bias and correct for the pixel window function, since the noise bias is removed in the catalogue-based estimator. The measured shear power spectrum for UNIONS data is shown in Fig.~\ref{fig:data_vector_and_best_fit}, along with the best-fitting model for our fiducial $\Lambda$CDM results described in Sect.~\ref{sec:results}.

In addition to the cosmological power spectrum, we compute the $EB$- and $BB$-mode power spectra; as $\alpha, \beta \in \{E, B\}$ in Eq.~\eqref{eq:estimate_power_spectrum},
these
components are natural outputs
of the measurement and are discussed in Sect.~\ref{sec:B_modes} and \cite{daleyUNIONSWeakLensing2026}.

\subsection{Modelling}
\label{sec:modelling}
This section presents the theoretical background for the observed shear power spectra and the different sources of systematic effects.
\subsubsection{Cosmic shear power spectrum}
\label{sec:cs_power_spectrum}
We compute the cosmic shear power spectrum using the Limber approximation \citep{limberAnalysisCountsExtragalactic1953}. This approximation is valid as long as we are not considering the largest scales \citep[small multipoles,][]{lemosEffectLimberFlatsky2017, kilbingerPrecisionCalculationsCosmic2017}. The angular power spectrum can be computed as
\begin{align}
\label{eq:cs_power_spectrum}
    C_\ell = \int_0^{\chi_\mathrm{H}} {\rm d}\chi \frac{q^2(\chi)}{\chi^2} P_\mathrm{NL}\left[k=\frac{\ell + 1/2}{\chi}, z(\chi) \right],
\end{align}
where $\chi$ is the comoving distance, $\chi_\mathrm{H}$ is the comoving horizon distance, and $P_\mathrm{NL}$ is the non-linear 3D matter power spectrum. The lensing efficiency is given by
\begin{align}
    q(\chi) = \frac 3 2 \Omega_{\rm m} \frac{H_0^2}{c^2} \frac{\chi}{a(\chi)} \int_\chi^{\chi_\mathrm{H}} {\rm d}\chi' n(\chi') \frac{\chi-\chi'}{\chi'},
\end{align}
where $\Omega_{\rm m}$ is the matter density parameter, $H_0$ is the Hubble constant, $a = 1/(1+z)$ is the scale factor, and $n(\chi)$ is the redshift distribution of source galaxies described in Sect.~\ref{sec:redshift distribution}.

To perform the modelling, the linear part of the power spectrum is computed using \texttt{CAMB} \citep{lewisEfficientComputationCosmic2000}. The linear power spectrum model takes five cosmological parameters as input, $\omega_{\rm m}=\Omega_{\rm m} h^2$, $H_0 = 100h$ km s$^{-1}$ Mpc$^{-1}$, the amplitude of the primordial power spectrum, $A_\mathrm{s}$, and its tilt, $n_\mathrm{s}$, and the baryon density $\omega_{\rm b} = \Omega_{\rm b} h^2$. Some of these parameters are derived from the sampled parameters described in Sect.~\ref{sec:inference_choices}.

The linear power spectrum is not a sufficient description of matter clustering at scales probed by cosmic shear. At small scales, the growth of structure is non-linear \citep{jainCosmologicalModelPredictions1997, bernardeauLargescaleStructureUniverse2002} and affected by baryonic feedback from supernovae and active galactic nuclei \citep{vandaalenEffectsGalaxyFormation2011,semboloniQuantifyingEffectBaryon2011, chisariImpactBaryonsMatter2018, schallerFlamingoProjectBaryon2025}. We model this non-linear part of the power spectrum using \texttt{HMCode2020} \citep{meadHmcode2020ImprovedModelling2021}, based on the halo model, fitting physically motivated parameters to $N$-body and hydrodynamical simulations. The code provides a version that models baryonic feedback with a single parameter, $T_\mathrm{AGN}$, that controls the feedback amplitude. We will use this model to obtain our predictions for the cosmic shear power spectrum. Calibrations with $N$-body simulations estimate the value of the AGN temperature to be around $T_\mathrm{AGN} \sim 10^{7.6}-10^{8}$ K. In the inference, we marginalise on $\log T_{\rm AGN}$ with a uniform prior informed by hydrodynamical simulations (see Sect.~\ref{sec:inference})
\subsubsection{Intrinsic alignment}
\label{sec:intrinsic_alignment}
In addition to baryonic physics, cosmic shear is sensitive to astrophysical systematic effects due to the intrinsic alignment (IA) of galaxies \citep[see][for a recent review]{chisariRisingTideIntrinsic2025}. Galaxies are extended objects and are therefore sensitive to the tidal field of the gravitational potential. Galaxies forming in the same overdensity, therefore, tend to align with each other rather than having randomly distributed shapes. The observed shear power spectrum is therefore polluted by an additional contribution from the correlation of intrinsic shapes $C_{\ell, \mathrm{II}}$. Additionally, galaxies at different distances along the same line of sight experiencing lensing or gravitational tidal interactions from the same large-scale structure will contribute to the cosmic shear power spectrum with an additional cross-correlation term, $C_{\ell, \mathrm{\gamma I}}$, such that
\begin{align}
    C_\ell^{\mathrm{obs}} = C_{\ell, \mathrm{\gamma \gamma}} + C_{\ell, \mathrm{\gamma I}} + C_{\ell, \mathrm{I\gamma }} + C_{\ell, \mathrm{II}},
\end{align}
where the angular power spectra can be expressed in terms of the 3D power spectra, assuming the Limber approximation:
\begin{align}
    C_{\ell, \mathrm{II}} &= \int_0^{\chi_\mathrm{H}} d\chi \frac{n^2(\chi)}{\chi^2}P_\mathrm{II}\left[k = \frac{\ell + 1/2}{\chi}, z(\chi) \right];\\
    C_{\ell, \mathrm{\gamma I}} &= \int_0^{\chi_\mathrm{H}} \frac{q(\chi)n(\chi)}{\chi} P_\mathrm{\gamma I} \left[ k=\frac{\ell +1/2}{\chi}, z(\chi)\right].
\end{align}

We modelled intrinsic alignment using the non-linear linear alignment (NLA) model \citep{hirataIntrinsicAlignmentlensingInterference2004, bridleDarkEnergyConstraints2007}. This model extends the linear alignment model to small scales, effectively replacing the linear matter power spectrum, $P_\mathrm{L}(k, z)$, by the non-linear one, $P_\mathrm{NL}(k, z)$. The power spectra used to model intrinsic alignment were then
\begin{align}
    P_\mathrm{II}(k,z) &= \left(\frac{C_1(z) \rho_\mathrm{crit} \Omega_{\rm m}}{\bar{D}}\right)^2 P_\mathrm{NL}(k,z);\\
    P_\mathrm{\gamma I}(k,z) &= - \frac{C_1(z) \rho_\mathrm{crit} \Omega_{\rm m}}{\bar{D}} P_\mathrm{NL}(k,z),
\end{align}
where 
\begin{align}
    C_1(z) = A_\mathrm{IA}\bar{C}_1\left( \frac{1+z}{1+z_0}\right)^{\eta_\mathrm{IA}}.
\end{align}
Here, $\bar{D}(z)\propto (1+z)D(z)$, with $D(z)$ is the growth factor normalised to unity today, $\bar{C}_1=5 \times 10^{-14} (h^2 M_\odot /$Mpc$^{3} )^{-1}$ is a normalisation constant \citep{hirataIntrinsicAlignmentlensingInterference2004}, and $z_0$ is a pivot redshift. Since our measurement is non-tomographic, we cannot break the degeneracy between intrinsic alignment and cosmic shear. We therefore removed the redshift dependence by setting $\eta_\mathrm{IA}=0$, and we placed a strong yet conservative prior on the amplitude of the intrinsic alignment power spectrum $A_\mathrm{IA}$ (see Sect.~\ref{sec:intrinsic_alignment_prior} for details on this procedure).
%
\subsection{Covariance modelling}
\label{sec:cov_modelling}
We assume that the cosmic shear power spectrum follows a multivariate Gaussian distribution with fixed covariance. We rely on a theoretical estimate of the covariance to perform the inference detailed in Sect.~\ref{sec:inference}.

The analytical expression of the covariance of the $E$-mode of the shear power spectrum can be written as a sum of Gaussian and non-Gaussian contributions from the shear field. The Gaussian contribution is computed with \texttt{NaMaster} using the improved narrow kernel approximation (iNKA) estimator \citep{garcia-garciaDisconnectedPseudoClCovariances2019, nicolaCosmicShearPower2021}. The estimator accounts for mode mixing due to masking and binning within the pseudo-$C_\ell$ framework described in Sect.~\ref{sec:pseudo_cl}. The code requires the mode-coupled pseudo-$C_\ell$ spectra computed for the theoretical full-sky spectra convolved by the mixing matrix introduced in Eq.~\eqref{eq:pseudo_cl}. We add the noise bias to the theoretical power spectrum, estimated using the following analytical expression derived in \cite{nicolaCosmicShearPower2021} for the binned noise pseudo-power spectrum:
\begin{align}
\tilde{N}_L = \Omega_\mathrm{pix} \left \langle \underset{i \in p}{\sum} w_i^2 \frac{e^2_{1,i} + e^2_{2,i}}{2}\right \rangle_p,
\end{align}
where $\Omega_\mathrm{pix}$ is the area of a pixel in steradians and the average is performed over all pixels. We work with a resolution of $N_\mathrm{side}=1024$ and weight the generated maps using the galaxy count maps\footnote{The resolution corresponds to a pixel size of about $3.4$ arcmin.}. This estimate is equivalent to the noise bias obtained by projecting the weak lensing samples onto the \texttt{HealPix} map after applying random rotations to their shapes. This process correctly accounts for the shape noise and the variation of the number density on the sky. A fundamental difference with the catalogue-based estimator used in Sect.~\ref{sec:pseudo_cl}, however, is the use of the pixelised shear field to estimate the covariance in the absence of a catalogue-based estimator for the covariance within \texttt{NaMaster}.

The non-Gaussian contribution to the covariance is the sum of the connected four-point covariance (cNG) arising from the shear field trispectrum, and the so-called supersample covariance (SSC), accounting for correlations of multipoles used in the analysis with supersurvey modes with larger wavelengths than the survey used \citep[see][for more details]{reischkeKiDSLegacyCovarianceValidation2025}. The non-Gaussian terms are computed using \texttt{OneCovariance}. This software also calculates the Gaussian part of the covariance described above.

\begin{figure}
    \centering
    \includegraphics[width=1\linewidth]{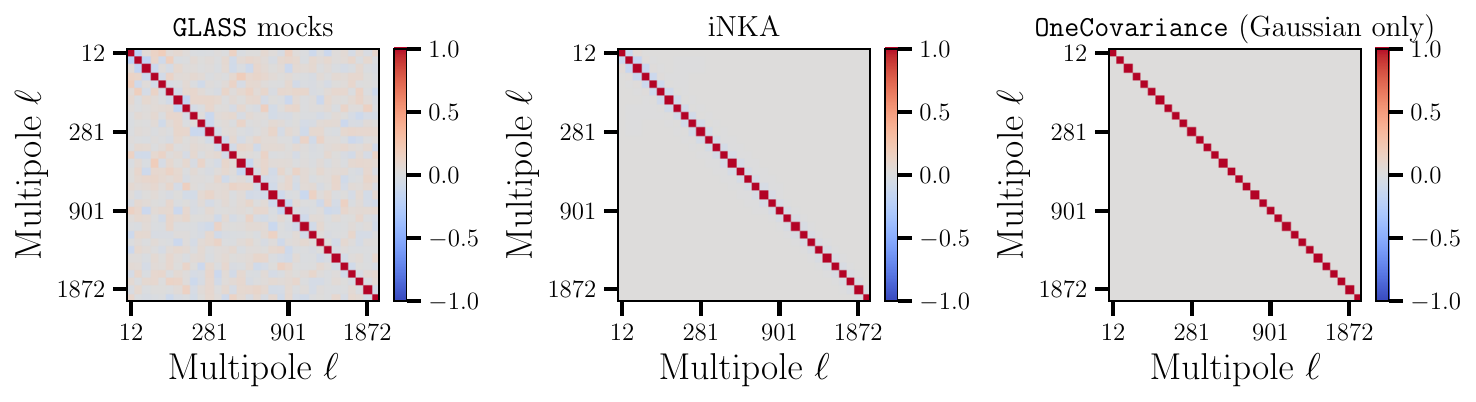}
    \includegraphics[width=1\linewidth]{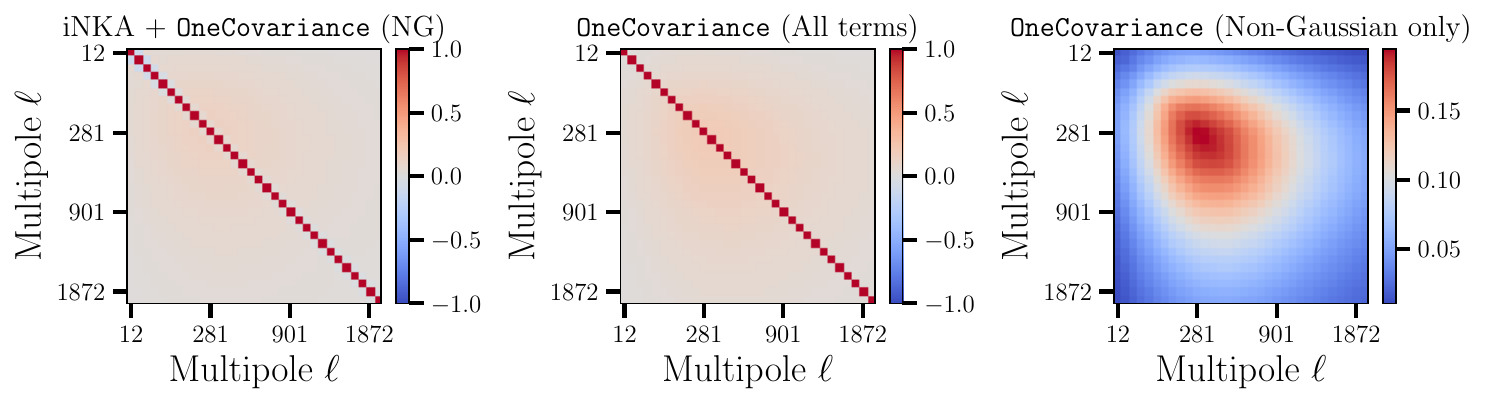}
    \caption{ Correlation matrix of the non-tomographic cosmic shear power spectrum. The covariance matrix is estimated using two theory prescriptions, iNKA and \texttt{OneCovariance}. It is compared to the covariance estimated from \texttt{GLASS} mocks. \textit{Top panels}: Gaussian parts of the theory covariance compared to the simulation covariance. \textit{Bottom}: Full covariance, including the non-Gaussian part of \texttt{OneCovariance}. The bottom right panel shows the non-Gaussian part of the covariance only, normalised with the full covariance diagonal.}
    \label{fig:corr_matrix}
\end{figure}

\begin{figure}
    \centering
    \includegraphics[width=1\linewidth]{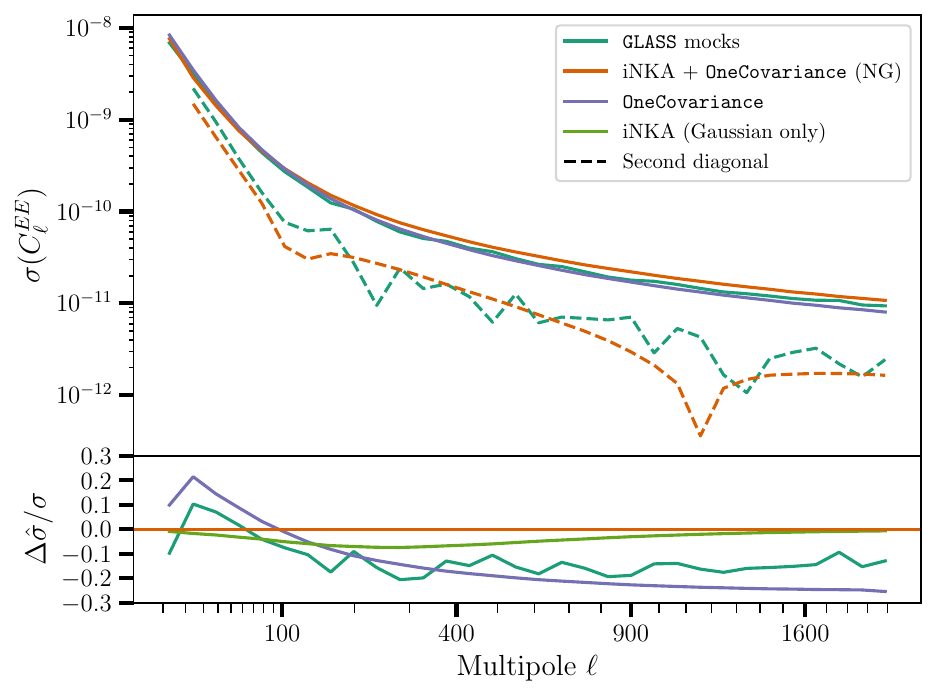}
    \caption{Comparison of the error bars of the theory prescriptions and the covariance estimated from \texttt{GLASS} mocks. \textit{Top panel}: error bars on the diagonal in solid lines. The dashed lines correspond to the second diagonal. \textit{Bottom}: Relative error compared to the fiducial covariance matrix used for the analysis, specifically the iNKA Gaussian part and the \texttt{OneCovariance} non-Gaussian part. There is a discrepancy of $15\%$ between the \texttt{NaMaster} estimate and the one obtained from mocks. Some of this discrepancy can be explained by a spatially varying number density (see Appendix~\ref{app:density_variation_covariance}).}
    \label{fig:errorbar_non_gaussian_cov}
\end{figure}

We validate the covariance using \texttt{OneCovariance} Gaussian and non-Gaussian part, and from 350 \texttt{GLASS} \citep{tessoreGLASSGeneratorLarge2023} log-normal mocks reproducing the footprint, the number density, the shape noise, and the redshift distribution of the UNIONS weak lensing sample. The fiducial cosmology used to generate the covariance is the \textit{Planck} 2018 cosmology \citep{planckcollaborationPlanck2018Results2020}. The correlation matrices obtained using these different methods are shown in Fig.~\ref{fig:corr_matrix}. The corresponding error bars are presented in Fig.~\ref{fig:errorbar_non_gaussian_cov}.

The lower right panel in Fig.~\ref{fig:corr_matrix} shows the correlation matrix of the non-Gaussian covariance only, normalised using the diagonal of the full covariance. The non-Gaussian terms add extra covariance at different multipoles, which becomes non-negligible for $100 \lesssim \ell \lesssim 900$, with correlations at the roughly $10\%$ level. This translates into an increase of the error bars on the diagonal of the covariance matrix of around $5\%$ at $\ell = 300$, as shown in the lower panel of Fig.~\ref{fig:errorbar_non_gaussian_cov}. The upper panel of Fig.~\ref{fig:errorbar_non_gaussian_cov} shows the error bars after adding the non-Gaussian contributions computed with \texttt{OneCovariance} to the Gaussian part of iNKA or \texttt{OneCovariance}. We find good agreement between the three methods on the largest scales ($\ell < 100$). On small scales, the error bars estimates obtained from \texttt{OneCovariance} and the \texttt{GLASS} mocks agree at the $5$ to $10\%$ level. However, the estimate computed from the Gaussian part of iNKA tends to overestimate the variance. We discuss in Appendix~\ref{app:density_variation_covariance} how this can be partly explained by a varying number density across the sky in the UNIONS survey. The lower panel shows the relative difference using iNKA combined with the \texttt{OneCovariance} non-Gaussian terms as the baseline. The discrepancy with \texttt{OneCovariance} goes up to about $20\%$. We adopt iNKA as fiducial because its noise-bias treatment cpatures the inhomogeneous galaxy density in UNIONS; \texttt{OneCovariance}'stationary-noise assumption ignores this contribution, which accounts for most of the about $20\%$ discrepancy on scales above $\ell$ of a few hundreds (see App.~\ref{app:density_variation_covariance}).  We check the impact of this difference on cosmological inference in Sect.~\ref{sec:robustness_modelling_covariance}.

The iNKA estimator introduced above also provides an estimate of the covariance matrix of the $B$-mode shear power spectrum and the cross-covariance between $E$- and $B$-mode power spectra. This estimator is used in Sect.~\ref{sec:B_modes} to assess the significance of the $B$ mode in the measured signal.

\section{Inference pipeline validation}
\label{sec:inference}

We perform our analysis using a multivariate Gaussian likelihood with fixed covariance within the Bayesian framework. Bayes' theorem \citep{bayesLIIEssaySolving1997} provides the probability distribution of the parameters $\theta$ given a model $\model$ and the observed data $d$. The probability distribution $p(\theta \,|d, \model)$ is given by
\begin{align}
    p(\theta \, |d, \model) = \frac{ \mathcal{L}(\theta) \times \Pi(\theta \,|\model)}{\mathcal{Z}(d \,|\model)},
\end{align}
where $\Pi(\theta \, |\model)$ is the prior distribution of the parameters, $\mathcal{Z}(d \, |\model)$ is the Bayesian evidence, which gives the probability of observing the data after marginalising over the parameters $\theta$, and $\mathcal{L}(\theta) \equiv p(d \, |\theta, \model)$ is the likelihood function. This corresponds to the probability of observing the data $d$ given the parameters $\theta$ (and the chosen model $\model$). In this work, we assume the following form of the likelihood:
\begin{align}
    -2 \log \mathcal{L} \propto [d-T(\theta)]^{\sf T} C^{-1} [d-T(\theta)],
\end{align}
where $d$ is the data vector. $T(\theta)$ is the prediction obtained from \texttt{CosmoSIS} \citep{zuntzCosmoSISModularCosmological2015} and $C$ is the covariance matrix of the data whose computation has been detailed in Sect.~\ref{sec:cov_modelling}. In Appendix~\ref{app:blinding}, we briefly describe the blinding procedure at the redshift distribution level adopted in this work.

\subsection{Inference choices}
\label{sec:inference_choices}

We use a sampling strategy similar to \cite{desandkidscollaborationY3KiDS1000Consistent2023}. We sample the cosmological parameters $\{\omega_{\rm c}, \omega_{\rm b}, H_0, n_\mathrm{s}, S_8\}$. The parameter estimation is performed using the \texttt{Polychord} \citep{handleyPolychordNestedSampling2015, handleyPolychordNextgenerationNested2015} nested sampler.

Our priors follow the Hybrid setup in \citet{desandkidscollaborationY3KiDS1000Consistent2023}. We impose additional priors on $\omega_{\rm b}$ and nuisance parameters, summarised in table 2 of \cite{gohUNIONSWeakLensing2026}\footnote{Note that the prior on PSF leakage parameters in \cite{gohUNIONSWeakLensing2026} is not used in this work (see Appendix~\ref{app:PSF_impact}).}.


\subsection{Intrinsic alignment prior}
\label{sec:intrinsic_alignment_prior}

Due to our non-tomographic measurement, we cannot efficiently break the degeneracy between intrinsic alignments and the amplitude of structure growth. To constrain $S_8$, we must therefore rely on a well-motivated prior on the amplitude of intrinsic alignment $A_{\rm IA}$. This prior can be built using the multiple direct intrinsic alignment measurements on early-type galaxies \citep{joachimiConstraintsIntrinsicAlignment2011, mandelbaumWiggleZDarkEnergy2011, singhIntrinsicAlignmentsSDSSIII2015, johnstonKiDS+GAMAIntrinsicAlignment2019, fortunaKiDS1000ConstraintsIntrinsic2021, liKiDS1000CosmologyImproved2023,hervaspetersUNIONSDirectMeasurement2025, navarro-gironesPAUSurveyMeasuring2025}. A similar approach has been used in previous cosmic shear analyses \citep{fortunaHaloModelVersatile2021,wrightKiDSLegacyCosmologicalConstraints2025}. Our weak lensing sample also includes late-type galaxies, which are expected to have a weaker intrinsic alignment signal. The combination of red and blue galaxy measurements gives us a prior for $A_{\rm IA} = 0.83 \pm 0.39$, the uncertainty of which we double when setting the prior. The procedure is detailed thoroughly in \cite{gohUNIONSWeakLensing2026}.
\subsection{Point spread function}
\label{sec:PSF}
Point spread function (PSF) systematics arise due to PSF mismodelling or when anisotropic PSF shapes leak into the measured shape of galaxies. These two effects can lead to both additive and multiplicative biases of the shear. The multiplicative bias is calibrated using \texttt{MetaCalibration} and image simulations \citep{hervaspetersUNIONSWeakLensing2026}. The additive bias can be estimated using galaxy-PSF correlation functions \citep{guerriniGalaxyPointSpread2025}. The use of correlation functions in configuration space, as presented in \cite{gohUNIONSWeakLensing2026}, has enabled the modelling of the PSF systematic effects and their incorporation in the inference pipeline. In Appendix~\ref{app:PSF_impact}, we check that the impact of PSF systematic effects on the power spectrum is small enough to be neglected in the modelling of the power spectrum and only rely on the catalogue-level calibration detailed in \cite{kilbingerUNIONSWeakLensing2026}. Section~\ref{sec:robustness_modelling_psf} checks the sensitivity of the results to the leakage calibration. 

%
%
\subsection{$B$-modes null test}
\label{sec:B_modes}
As mentioned in Sect.~\ref{sec:pseudo_cl}, gravitational lensing does not produce $B$ modes, to first order in the shear field and under the Born approximation. The $B$ modes produced by source clustering, intrinsic alignments, or other higher-order effects are expected to be small compared to the noise level in UNIONS data. However, as discussed in Sect.~\ref{sec:PSF}, systematic effects like PSF leakage can produce larger $B$-modes in practice. Figure~\ref{fig:tau_0_cell} shows that some $B$-modes due to PSF systematics might pollute the cosmic shear signal. Measuring $B$-mode significance is therefore informative but not sufficient as a test for systematic effects.

\begin{figure}
    \centering
    \includegraphics[width=1\linewidth]{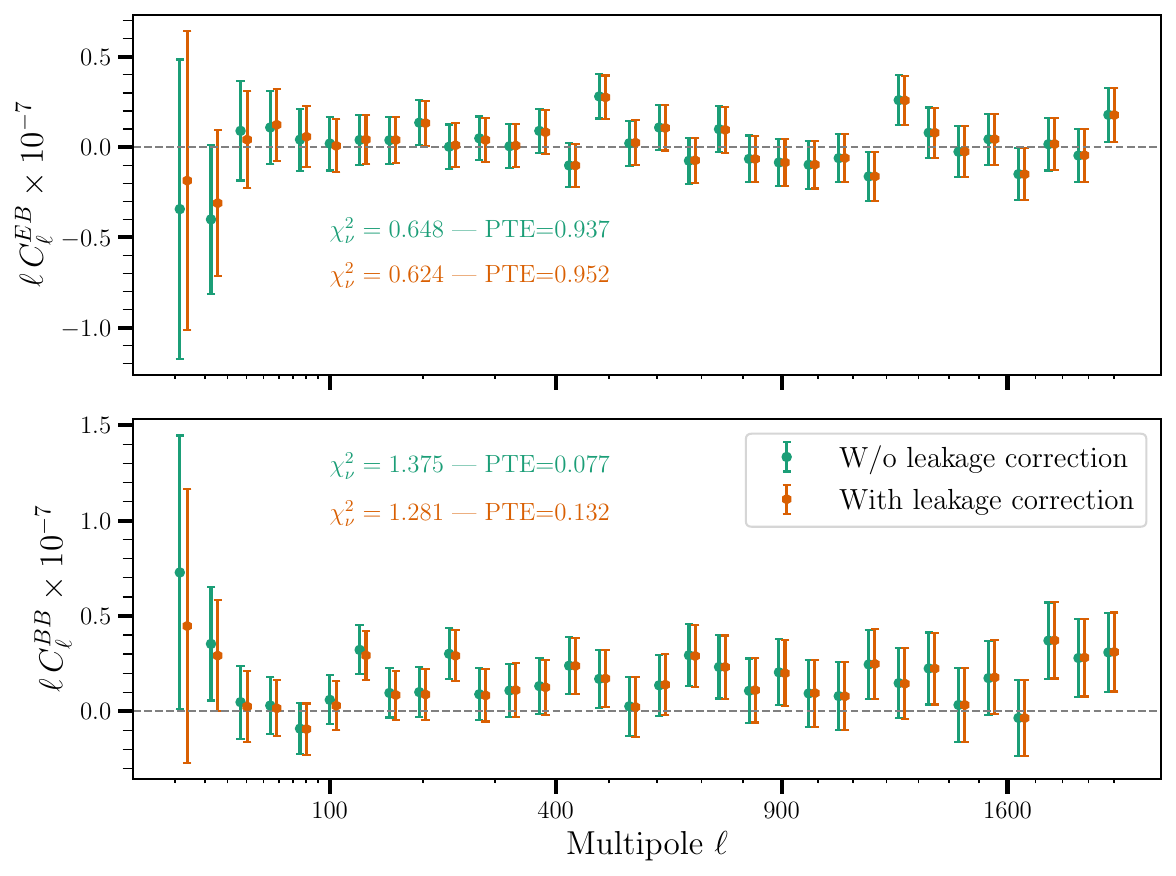}
    \caption{ $EB$ and $BB$ power spectra measured on the non-tomographic UNIONS data. Data points for the leakage-corrected case are slightly offset for visualisation purposes. The covariance is estimated using iNKA (see Sect.~\ref{sec:cov_modelling} for details) and validated against Gaussian simulations. The $\chi^2$ and probability-to-exceed (PTE) are reported in the figure. When considering all scales, our findings are consistent with the null hypothesis of no $B$ modes. We observe that removing leakage slightly improves the PTE; however, the test marginally passes without scale cuts. Removing the three largest multipoles ($\ell > 1700$) further improves the PTEs of the $BB$-mode null test to $0.12$ and $0.21$ for the catalogues with and without leakage correction, respectively.}
    \label{fig:eb_bb_power_spectrum}
\end{figure}

Figure~\ref{fig:eb_bb_power_spectrum} shows the $EB$ and $BB$ power spectra measured on the non-tomographic UNIONS data, using the procedure described in Sect.~\ref{sec:pseudo_cl}. The covariance is estimated using iNKA (see Sect.~\ref{sec:cov_modelling} for details). For the $EB$ power spectrum, we obtain a $\chi^2$ of $20$ for $32$ degrees of freedom, leading to a probability-to-exceed (PTE) of $0.95$. For the $BB$ power spectrum, we measure a $\chi^2$ of $41$ for $32$ degrees of freedom, amounting to a PTE of $0.136$. We set our PTE test threshold to $0.05$. In \citet{daleyUNIONSWeakLensing2026}, $B$ modes in configuration space are identified at both small and large scales. While the null test marginally passes, the $BB$ spectrum shows a small residual amplitude. Removing the three points at the smallest scales improves the $B$-mode null test, increasing the PTE to $0.231$, a $70\%$ increase. In addition, removing points below $\ell_\mathrm{min}=300$ further improves the PTE to $0.277$. \cite{daleyUNIONSWeakLensing2026} provides further details on the $B$-mode measurements in configuration and harmonic space, compares both estimators, and motivates the scale cuts used in configuration space and in this work (see Sect.~\ref{sec:scale_cut}). The covariance used for the $B$-mode null test is validated against Gaussian simulations in Appendix~\ref{app:B_mode_cov_validation} following the procedure described in \cite{douxDarkEnergySurvey2022}.

\subsection{Scale cuts}
\label{sec:scale_cut}
Modelling uncertainties at small scales in the cosmic shear power spectrum and unmodelled systematic effects can bias cosmological inference if not properly accounted for. A strategy to mitigate the bias is to cut the scales where the model cannot accurately capture the physical effects or where unknown systematic effects dominate the signal. In this work, we motivate our choice of scale cut from both theoretical considerations and $B$-mode measurements. The scale cuts to mitigate the sensitivity to baryonic feedback are discussed in Appendix~\ref{app:scale_cut}. We derive that to limit the sensitivity of our analysis to physical scales above $k_{\rm max} = 3 \, h \, \textrm{Mpc}^{-1}$.

In addition, based on the PTE presented in \cite{daleyUNIONSWeakLensing2026} and Sect.~\ref{sec:B_modes}, we remove the largest and smallest scales, where a potential $B$ mode signal is present. To be conservative, we use $\ell_\mathrm{min}=300$ and $\ell_\mathrm{max}=1600$. This choice of $\ell_\mathrm{max}$ corresponds to a maximum wavenumber $k_\mathrm{max} \thickapprox 2.6 h \, {\rm Mpc}^{-1}$ in the framework used in Appendix~\ref{app:scale_cut}. This choice of scale cuts becomes our fiducial setup and is used in Sect.~\ref{sec:fiducial_analysis}. We show in Appendix~\ref{app:B_mode_cov_validation} that this choice of cuts is robust to the choice of covariance estimator used in the $B$-mode null test.

For consistency, we run inference by splitting the data vector into small and large scales. For robustness in modelling, we use the methodology above for $k_\mathrm{max} \in [1, 3, 5] h \, {\rm Mpc}^{-1}$, while keeping our scale cut on large scales fixed at $\ell_\mathrm{min}=300$. We also run our inference with \texttt{Halofit} and \texttt{HMCode} without baryons to test the sensitivity to baryonic feedback modelling. Finally, we relax the large-scale cut while keeping our fiducial small-scale cut, $\ell_\mathrm{max}=1600$, to check the impact of large-scale $B$ modes on our results. Results are discussed in Sect.~\ref{sec:robustness_modelling_sc}.
\subsection{Redshift distribution uncertainties}
\label{sec:redshift_uncertainty}
We include uncertainties on the photometric redshifts by allowing overall translations of the fiducial redshift distribution, shown in Fig.~\ref{fig:nz}, such that
\begin{align}
    n(z) \rightarrow n(z+\Delta z).
\end{align}
Redshift uncertainties are estimated using realistic mock catalogues \citep[see][and Sect. 4.4, for full details]{gohUNIONSWeakLensing2026} constructed from the MICE2 simulation \citep{fosalbaMICEGrandChallenge2015}. These mocks are designed to reproduce the photometric properties and selection functions of the UNIONS analysis, including CFHTLenS-like noisy $ugriz$ photometry \citep{hildebrandtCFHTLenSImprovingQuality2012}. We apply a noise-modelling framework inspired by \citet{vandenbuschTestingKiDSCrosscorrelation2020} to generate fluxes that reflect depth variations, seeing, and galaxy size. To ensure realistic selection functions, we employ the kNN-matching scheme of \citet{wrightKiDSLegacyRedshiftDistributions2025}, reproducing both the UNIONS--CFHTLenS matched photometric sample and the spectroscopic calibration sets including DEEP2 \citep{newmanDEEP2GalaxyRedshift2013}, VVDS \citep{lefevreVIMOSVLTDeep2005}, and VIPERS \citep{scodeggioVIMOSPublicExtragalactic2018}.

The mock photometric and spectroscopic samples are then passed through the same SOM-based redshift-calibration pipeline used in the configuration space analysis (see Sect.\,\ref{sec:redshift distribution}), using the shape and shear-response weights assigned by the kNN matching. In the mocks, we compare the SOM-recovered redshift distribution to the true MICE2 redshifts, yielding an estimate of the systematic bias in the mean redshift, $\Delta z = -0.003$.

To quantify the uncertainty on the bias, we use the standard deviation of bootstrap realisations of the SOM-based $n(z)$ from the real data. This provides a data-driven estimate of the statistical uncertainty, which we verify to be consistent with the scatter measured in the mock catalogues. Adding a systematic uncertainty, we adopt a total uncertainty of $\pm 0.018$ as the width of the prior on the mean redshift shift parameter.

\subsection{Shear multiplicative bias}
\label{sec:multiplicative_bias}
The bulk of the shear multiplicative bias is corrected for by \texttt{MetaCalibration}. However, residual multiplicative bias can remain due to blending, which is not accurately corrected for by \texttt{MetaCalibration}. The residual multiplicative bias and its uncertainty are estimated from image simulations \citep{hervaspetersUNIONSWeakLensing2026}. The residual multiplicative bias $m$ is then added to the inference as a nuisance parameter and rescales the power spectrum, such that
\begin{align}
    C_\ell \rightarrow (1+m)^2 C_\ell.
\end{align}
%
\section{Results}
\label{sec:results}

This section presents the main result of the paper, explores the robustness to modelling choices, assesses the consistency with configuration space results from \cite{gohUNIONSWeakLensing2026}, and compares to other weak lensing experiments.

\subsection{Fiducial analysis}
\label{sec:fiducial_analysis}
\begin{figure}
    \centering
    \includegraphics[width=1\linewidth]{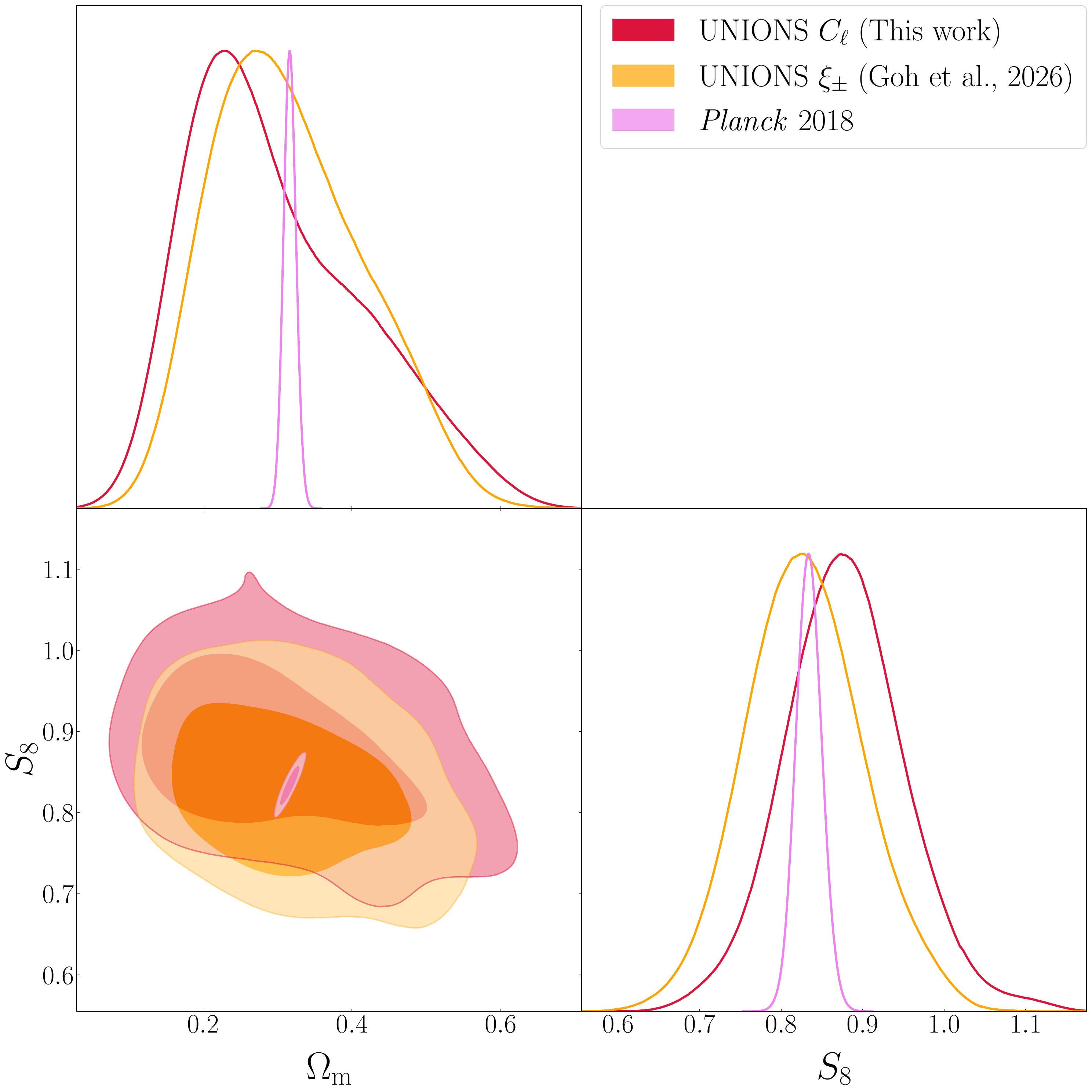}
    \caption{1D and 2D posteriors on $S_8 \equiv \sigma_8 (\Omega_{\rm m} / 0.3)^{0.5}$ and $\Omega_{\rm m}$ obtained using our fiducial analysis setup in harmonic space (this work, blue contours), configuration space \citep[][orange contours]{gohUNIONSWeakLensing2026}. This is compared to constraints from \textit{Planck} (pink contours). The consistency between configuration and harmonic space constraints on $S_8$ is discussed in Sect.~\ref{sec:consistency}. Our results are consistent with \textit{Planck} at the $0.79\, \sigma$ level.}
    \label{fig:cl_vs_xi}
\end{figure}

In this section, we present our constraints on $\Lambda$CDM assuming our fiducial model presented in Sects.~\ref{sec:methods} and~\ref{sec:inference}. In this fiducial setup, PSF-leakage-corrected ellipticities are used to reduce PSF systematics. We use NLA for intrinsic alignments without redshift dependence ($\eta_\mathrm{IA} = 0$). The scale cuts $\ell_{\rm min}=300$ and $\ell_\mathrm{max} = 1600$ are derived from Sect.~\ref{sec:scale_cut}, informed by $B$-mode analysis in Sect.~\ref{sec:B_modes} and \cite{daleyUNIONSWeakLensing2026}. These cuts correspond to $k_{\rm max} \approx 2.6 h \, {\rm Mpc}^{-1}$. The object-wise PSF leakage-corrected ellipticities are used to reduce PSF systematics contributions. The sampled parameters and their priors are summarised in Table 2 of \cite{gohUNIONSWeakLensing2026}. Figure~\ref{fig:cl_vs_xi} shows constraints obtained from this work, which are compared to the configuration-space constraints from \cite{gohUNIONSWeakLensing2026}. Constraints from \textit{Planck} are shown in pink.

Using the non-tomographic cosmic shear power spectrum, we find
\begin{align*}
    \Omega_{\rm m} &= 0.225_{-0.077}^{+0.153}, \text{[$C_\ell$, $\ell_\mathrm{min}=300$, $\ell_\mathrm{max}=1600$]},\\
    \sigma_8 &= 0.960_{-0.257}^{+0.223},\\
    S_8 &= 0.891_{-0.084}^{+0.057},\\
\end{align*}
where we report the maximum a posteriori and 68 per cent confidence intervals of the posterior computed using \texttt{getdist} \citep{lewisGetDistPythonPackage2025}. The corresponding theoretical shear power spectrum is shown in Fig.~\ref{fig:data_vector_and_best_fit} as a red solid line. The agreement of the data is good with a $\chi^2 = 8.1$, corresponding to a PTE of $0.76$ with respect to \texttt{GLASS} mocks (see Fig.~\ref{fig:chi2_glass_mock}). Our constraints are consistent with \textit{Planck} at the $0.79\, \sigma$ level, as shown in Fig.~\ref{fig:cl_vs_xi}. The constraints on $S_8$ are consistent with the configuration space analysis \citep{gohUNIONSWeakLensing2026} at the $0.60 \, \sigma$ level, assuming independent error bars. The consistency between the two analyses, accounting for the correlation between the statistics, is discussed in Sect.~\ref{sec:consistency}.

\begin{figure*}
    \centering
    \includegraphics[width=1\linewidth]{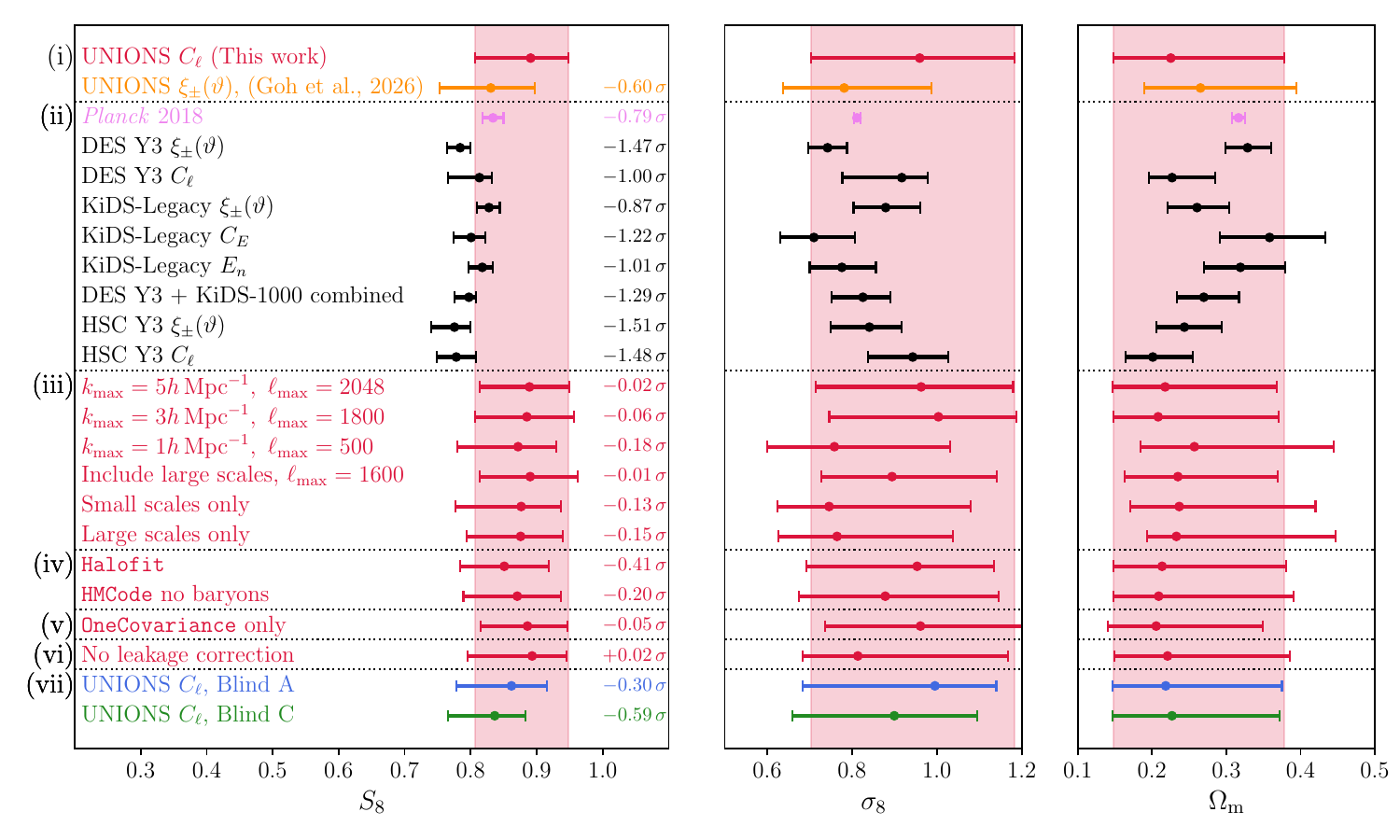}
    \caption{Comparison of 1D marginal posterior distributions over the parameters $S_8 \equiv \sigma_8 (\Omega_{\rm m} / 0.3)^{0.5}$, $\sigma_8$, and $\Omega_{\rm m}$ from UNIONS (this work), other experiments and consistency tests. (i) Constraints obtained from the harmonic space (this work) and configuration space \citep{gohUNIONSWeakLensing2026} analyses of UNIONS data are shown in red and orange, respectively. (ii) Comparison with constraints obtained from other experiments, including the weak lensing experiment in black. Our results are consistent with Planck at the $0.79\, \sigma$ level and with other weak lensing experiments at the $0.87$ to $1.51 \, \sigma$ level. (iii) Constraints obtained from varying the scale cut in harmonic space are shown for $k_\mathrm{max} \in [1, 3, 5] h \, {\rm Mpc}^{-1}$ (see Sect.~\ref{sec:scale_cut}) and keeping only small or large scales. (iv) Constraints obtained by varying the non-linear matter power spectrum modelling using \texttt{Halofit} or \texttt{HMCode2020} without baryonic feedback. (v) Constraints obtained using both the Gaussian and non-Gaussian terms computed with \texttt{OneCovariance} to estimate the covariance matrix. (vi) Comparison of the constraints obtained without applying the empirical leakage calibration (see Sect.~\ref{sec:PSF}). (vii) Comparison of the constraints obtained with the two false redshift blinds (see App.~\ref{app:blinding}).}
    \label{fig:whisker_plot}
\end{figure*}

\begin{figure}
    \centering
    \includegraphics[width=1\linewidth]{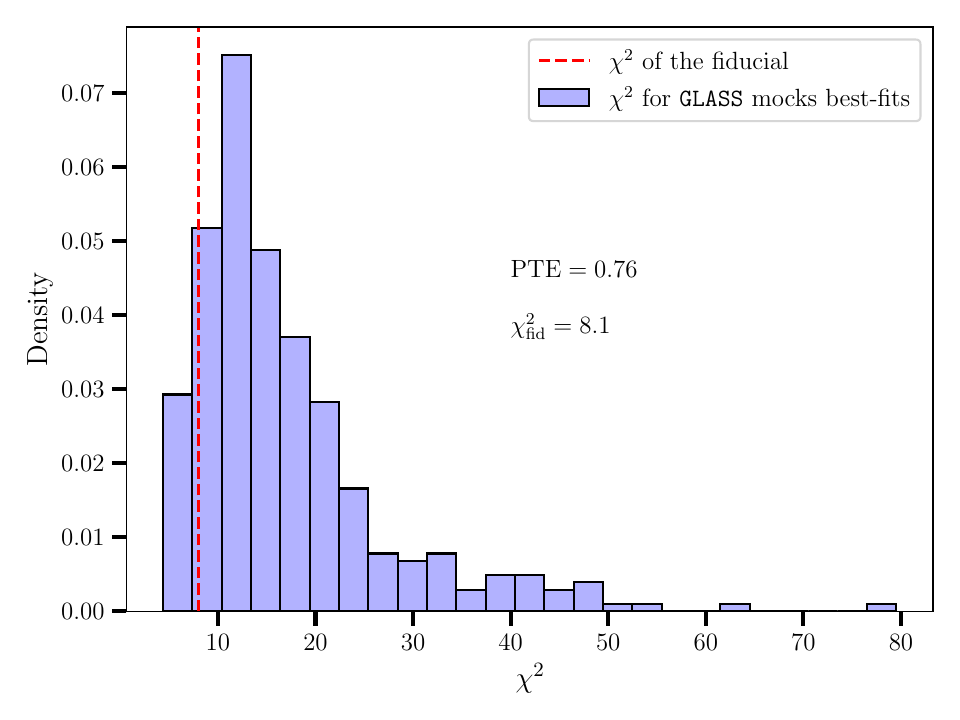}
    \caption{Distribution of $\chi^2$ values obtained on the \texttt{GLASS} mocks at each evaluated best-fit. The vertical dashed line shows the $\chi^2$ value obtained on the data for our fiducial analysis setup. The PTE is computed as the fraction of mocks with a $\chi^2$ larger than the one obtained on the data.}
    \label{fig:chi2_glass_mock}
\end{figure}

The 1D marginal constraints are also shown in Fig.~\ref{fig:whisker_plot} along with constraints for variations of the analysis further developed in Sect.~\ref{sec:robustness_modelling} and comparisons to other experiments discussed below. Appendix~\ref{app:additional_results} shows the full posterior for our fiducial setup in Figure~\ref{fig:full_posterior}, summarises results and metrics in Table~\ref{tab:metrics} and provides additional discussion on the obtained constraints. The choice of estimator to report the point estimate is discussed in Appendix~\ref{app:point_estimate}.


\subsection{Robustness to modelling choices}
\label{sec:robustness_modelling}

We assess the robustness of the cosmological constraints against analysis choices in Sects.~\ref{sec:methods} and \ref{sec:inference}.

\subsubsection{Scale cuts}
\label{sec:robustness_modelling_sc}

Scale cuts are derived in Sect.~\ref{sec:scale_cut} using \texttt{HMCode2020} non-linear power spectrum modelling and baryonic feedback emulation. Given a $k_\mathrm{max}$, an upper limit multipole $\ell_\mathrm{max}$ is derived such that the angular power spectrum depends by less than 5\% on $k$-scales larger than $k_\mathrm{max}$. To test the sensitivity of our analysis to scale cuts, we run inference using $k_\mathrm{max} = 5\,h\,\mathrm{Mpc}^{-1}$ ($\ell_\mathrm{max}=2048$), $k_\mathrm{max} = 3\,h\,\mathrm{Mpc}^{-1}$ ($\ell_\mathrm{max}=1800$) and $k_\mathrm{max} = 1\,h\,\mathrm{Mpc}^{-1}$ ($\ell_\mathrm{max} = 500$) while keeping the lower multipole limit to $\ell_\mathrm{min}=300$. Adding smaller scales gives consistent constraints with our fiducial setup ($-0.06 \, \sigma$ and $-0.02 \, \sigma$). Cutting at $\ell_\mathrm{max}=500$ shifts $S_8$ to a lower value by $-0.18 \, \sigma$. Including large scales (below $\ell_\mathrm{min}=300$) finds the same $S_8$ as the fiducial.

We also compare the constraints obtained with only small or only large scales. The cut is performed at $\ell_\mathrm{mid} = 800$ so that each inference runs on the same number of data points. In both cases the shift in $S_8$ is about $-0.14 \, \sigma$. Overall, we find that our results are robust to the choice of scale cuts, given a non-linear model for the matter power spectrum. 



\subsubsection{Point spread function leakage}
\label{sec:robustness_modelling_psf}

Our fiducial analysis uses PSF-leakage-corrected ellipticities \citep[see Sect.~\ref{app:PSF_impact} and][for details]{kilbingerUNIONSWeakLensing2026}. In App.~\ref{app:PSF_impact}, the additive bias due to PSF systematics is found to be negligible (see Fig.~\ref{fig:ratio_cell_cell_sys}). As a further check, we run an inference chain on the cosmic shear power spectrum measured using the uncorrected ellipticities. The leakage correction affects the power spectrum measurement, particularly on large scales, where leakage is most significant. The constraints with and without leakage correction are in agreement ($+0.02 \, \sigma$), showing the minimal impact of PSF systematics on our cosmological inference. In \cite{gohUNIONSWeakLensing2026}, a similar test is performed resulting in a $-0.13 \, \sigma$ shift on $S_8$ with respect to the fiducial analysis in configuration space, highlighting the reduced sensitivity to PSF leakage in harmonic space.


\subsubsection{Covariance modelling}

\label{sec:robustness_modelling_covariance}
In Sect.~\ref{sec:cov_modelling}, we identified a difference of around $25\%$ between the covariance estimated with iNKA and the one computed with \texttt{OneCovariance}. We ran our fiducial analysis replacing the iNKA covariance with the \texttt{OneCovariance} one. Using \texttt{OneCovariance} shifts $S_8$ to higher values by $0.02 \, \sigma$. It shows that the difference between the covariance discussed in App.~\ref{app:density_variation_covariance} does not impact the inference.


\subsubsection{Non-linear matter power spectrum modelling}

We run two additional inference chains varying the non-linear matter power spectrum modelling. This is not a test of sensitivity to dark matter models beyond $\Lambda$CDM, but of sensitivity to our choice of non-linear description, whose default is the \texttt{HMCode2020} including baryonic feedback as described in Sect.~\ref{sec:cs_power_spectrum}. For this test, we first replace \texttt{HMCode2020} with \texttt{Halofit} \citep{takahashiREVISINGHALOFITMODEL2012} to model the non-linear part of the matter power spectrum. Second, we use \texttt{HMCode2020} without baryonic feedback. Using \texttt{Halofit} and ignoring baryonic feedback in \texttt{HMCode2020} shifts $S_8$ to lower values by, respectively, $-0.41 \, \sigma$ and $-0.20 \, \sigma$.The observed shift is consistent with results reported in Figure 9 of \cite{guMitigatingNonlinearSystematics2025}. This shift is insignificant with our current 2D constraining power, but needs to be revisited for UNIONS tomographic analyses and upcoming Stage-IV analyses.


\subsection{Consistency with configuration space analysis}
\label{sec:consistency}

The harmonic space and configuration space analysis \citep{gohUNIONSWeakLensing2026} are consistent at the $0.60 \, \sigma$ level on $S_8$, assuming independent error bars (see Fig.~\ref{fig:whisker_plot}). However, this assumption is incorrect, since constraints are derived from the same data set. To quantify the consistency more accurately, we estimate the correlation between the two measurements using the \texttt{GLASS} mocks presented in Sect.~\ref{sec:cov_modelling}. For each of the $350$ realisations, we measure both the cosmic shear power spectrum and the two-point correlation functions. We then run a cosmological inference on each of these measurements using the same setup as for our fiducial analysis. This procedure yields $350$ samples of the cosmological parameters for both harmonic and configuration-space analyses.

\begin{figure}
    \centering
    \includegraphics[width=1\linewidth]{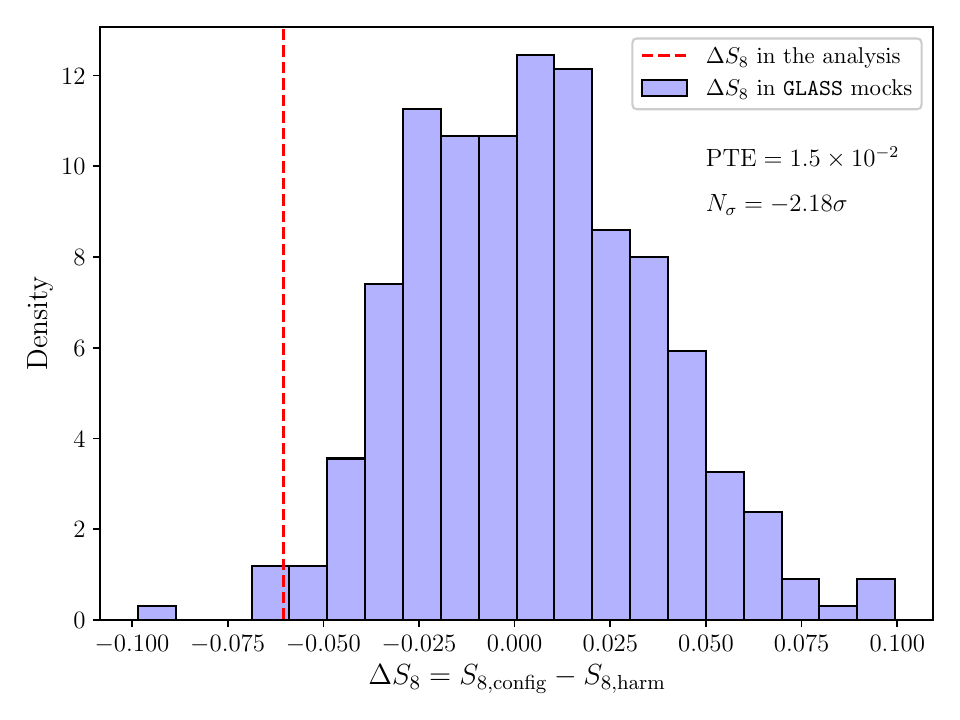}
    \caption{Histogram of the difference in estimated $S_8$ using the two-point correlation function and the power spectrum from the $350$ \texttt{GLASS} mocks. The difference measured with UNIONS is indicated with the red, dashed line.}
    \label{fig:difference_config_harm}
\end{figure}

Figure~\ref{fig:difference_config_harm} shows a histogram of the difference between the estimate of $S_8$ in configuration and harmonic space. The difference distribution is centred at around zero, but the observed difference of $0.06$ on $S_8$ has a $p$-value of $1.5 \times 10^{-2}$, which corresponds to a $2.18 \, \sigma$ difference. This estimate uncertainty is of about $\pm 0.2 \, \sigma$, treating the simulations count in the distribution tail as a Poissonian random variable.

We highlight that the different choices of scale cuts in the configuration and harmonic space do not alone explain the difference observed on $S_8$ between the two statistics. Indeed, Fig.~\ref{fig:whisker_plot} shows that the harmonic-space analysis is largely insensitive to the choice of scale cuts. Additionally, the large-small scales split does not reveal a significant difference on $S_8$ between the two scale ranges considered. However, $S_8$ value strongly depends on scale cuts in configuration space and including scales down to $5\arcmin$ brings its value at the level of the harmonic space constraints. The distribution of the localised $B$-mode in configuration space across multipoles $\ell$ is a reasonable candidate to explain the higher value of $S_8$ and its independence to scale cut at the same time. This will be further explored in future release together with the impact of upcoming shape measurement and sample selection refinements \citep{kilbingerUNIONSWeakLensing2026,hervaspetersUNIONSWeakLensing2026}.

\subsection{Comparison to other weak lensing surveys}
\label{sec:comparison_weak_lensing_survey}

\begin{figure}
    \centering
    \includegraphics[width=1\linewidth]{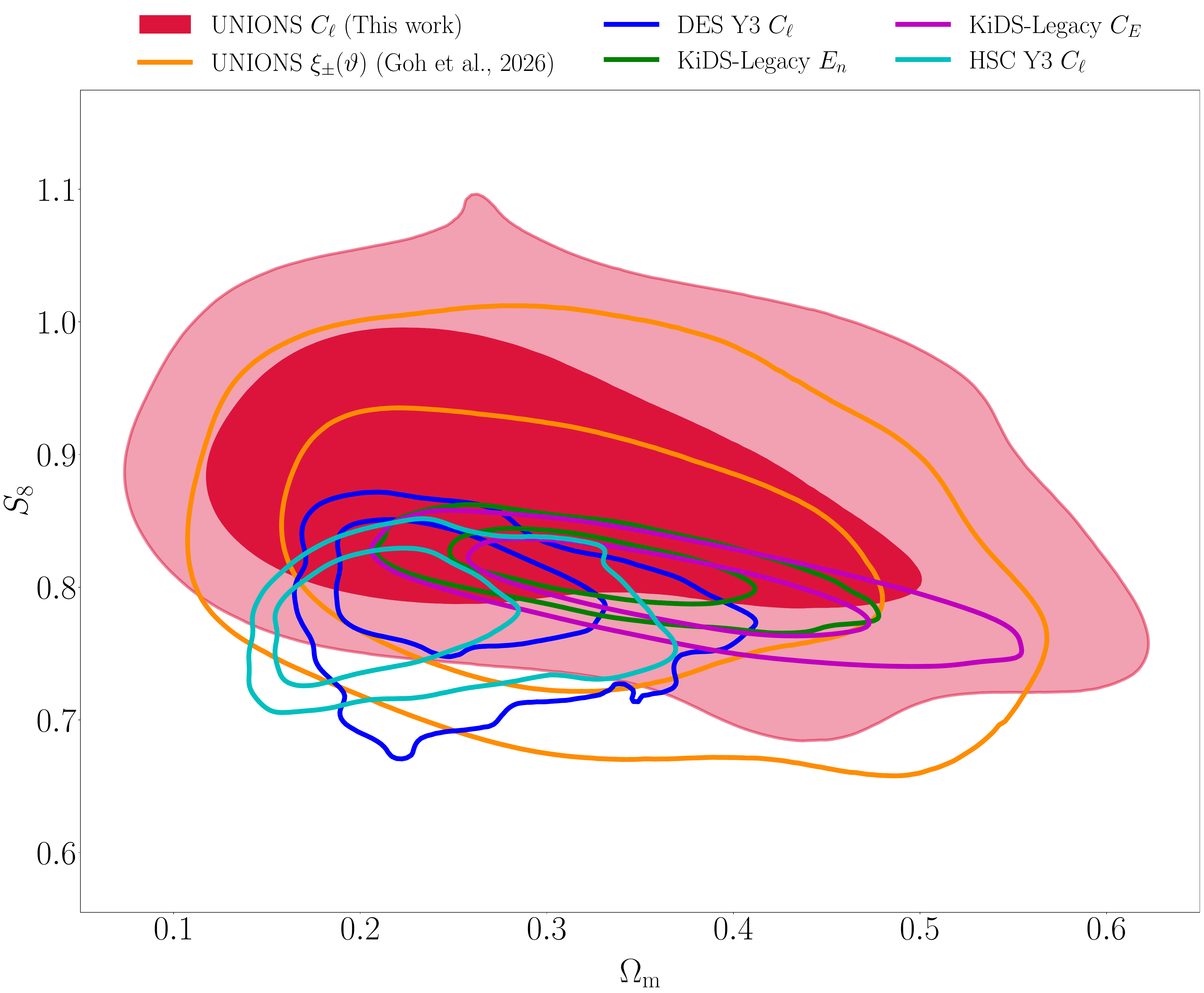}
    \caption{Comparison of 2D marginal posteriors on $S_8$ and $\Omega_{\rm m}$ obtained from UNIONS (this work, blue contour), DES Y3, HSC Y3, and KiDS-1000. Our results are consistent with other weak lensing experiments between $0.87$ and $1.51 \, \sigma$ level.}
    \label{fig:comparison_weak_lensing_surveys}
\end{figure}

Figure~\ref{fig:comparison_weak_lensing_surveys} compares the constraints described above with results from DES Y3 \citep{douxDarkEnergySurvey2022}, HSC Y3 \citep{dalalHyperSuprimeCamYear2023} and KiDS-Legacy \citep{wrightKiDSLegacyCosmologicalConstraints2025} in the $(\Omega_{\rm m}, S_8)$ plane. Our results are consistent with other weak lensing surveys at a level of $0.87$ to $1.51 \, \sigma$. The point estimate of $S_8$ obtained from UNIONS is notably large compared to other weak lensing surveys, but the non-tomographic setup and our conservative choices yield large error bars such that our results are consistent with both other weak lensing experiments and CMB. Our current caveat is a strong degeneracy between $S_8$ and the amplitude of intrinsic alignment, $A_{\rm IA}$, which is not well constrained by our non-tomographic analysis. We are confident that future work with tomographic binning will help calibrate the intrinsic alignment amplitude and break this degeneracy, allowing us to provide more insight into the $S_8$ tension.

\section{Conclusion}
In this work, we have used data from the Ultraviolet Near-Infrared Optical Northern Survey (UNIONS) to measure the non-tomographic cosmic shear power spectrum and constrain cosmological parameters within the $\Lambda$CDM framework. Our weak lensing sample, containing millions of galaxies, is obtained using \texttt{MetaCalibration} on the $r$ band data of the Canada-France Imaging Survey (CFIS) \citep[see][]{hervaspetersUNIONSWeakLensing2026}. We measure the cosmic shear power spectrum using a pseudo-$C_\ell$ approach \citep{alonsoUnifiedPseudoClFramework2019} that accounts for the complex survey geometry. The analysis is accompanied by a configuration-space analysis \citep{gohUNIONSWeakLensing2026}, allowing for cross-validation of the results.

We derived constraints on cosmological parameters using cosmic shear alone, following choices developed in \cite{desandkidscollaborationY3KiDS1000Consistent2023}. Using our non-tomographic setup, we find $S_8 = 0.891_{-0.084}^{+0.057}$ and $\Omega_{\rm m} = 0.225_{-0.077}^{+0.153}$. Our results are consistent with \textit{Planck} at the $0.79\, \sigma$ level and with other weak lensing surveys at the $0.87$ to $1.51 \, \sigma$ level. We assess the consistency of our harmonic space analysis with the configuration space analysis of \cite{gohUNIONSWeakLensing2026} and find a $2.18 \, \sigma$ difference on $S_8$ between the two analyses, accounting for the correlation between the two measurements. We also test the robustness of our results to modelling choices, including scale cuts, non-linear matter power spectrum modelling, and PSF systematics. We find that our results are robust to these choices. The strong correlation between the growth of structures and the amplitude of the intrinsic alignment remains a major caveat of our non-tomographic analysis, which will be addressed in future work with tomographic binning.

This work is part of the first cosmological analysis of UNIONS data. It demonstrates the potential of UNIONS for weak lensing cosmology and paves the way for future analyses with tomographic binning and additional statistics, which will allow us to provide more insight into the $S_8$ tension between weak lensing and CMB experiments. In particular, the unique overlap of UNIONS with spectroscopic surveys in the northern hemisphere opens exciting scientific opportunities for cross-correlation analyses, which will be explored in future work.

The codes used to measure the statistics, estimate the covariance matrix, and run the inference are publicly available\footnote{\nolinkurl{https://github.com/CosmoStat/sp_validation}}.

\section*{Data availability}
A subset of the raw data underlying this article is publicly available via the Canadian Astronomical Data Centre at \url{http://www.cadc-ccda.hia-iha.nrc-cnrc.gc.ca/en/megapipe/}.
The remaining raw data and all processed data are available to members of the Canadian and French communities via reasonable requests to the principal investigators of the Canada-France Imaging Survey, Alan McConnachie and Jean-Charles Cuillandre.

\begin{acknowledgements}
We thank Axel Guinot and Douglas Scott for constructive comments. We thank Axel Guinot for contributions throughout the different stages of this paper's elaboration. We thank Samuel Farrens for his contributions in developing and maintaining the \texttt{ShapePipe} library.
We are honoured and grateful for the opportunity of observing the Universe from Maunakea and Haleakala, which both have cultural, historical and natural significance in Hawai'i. This work is based on data obtained as part of the Canada-France Imaging Survey, a CFHT large program of the National Research Council of Canada and the French Centre National de la Recherche Scientifique. Based on observations obtained with MegaPrime/MegaCam, a joint project of CFHT and CEA Saclay, at the Canada-France-Hawai'i Telescope (CFHT) which is operated by the National Research Council (NRC) of Canada, the Institut National des Science de l’Univers (INSU) of the Centre National de la Recherche Scientifique (CNRS) of France, and the University of Hawai'i. This research used the facilities of the Canadian Astronomy Data Centre operated by the National Research Council of Canada with the support of the Canadian Space Agency. This research is based in part on data collected at Subaru Telescope, which is operated by the National Astronomical Observatory of Japan.
Pan-STARRS is a project of the Institute for Astronomy of the University of Hawai'i, and is supported by the NASA SSO Near Earth Observation Program under grants 80NSSC18K0971, NNX14AM74G, NNX12AR65G, NNX13AQ47G, NNX08AR22G, 80NSSC21K1572 and by the State of Hawai'i.
This work was made possible by utilising the CANDIDE cluster at the Institut d’Astrophysique de Paris. The cluster was funded through grants from the PNCG, CNES, DIM-ACAV, the Euclid Consortium, and the Danish National Research Foundation Cosmic Dawn Center (DNRF140). It is maintained by Stephane Rouberol. We gratefully acknowledge support from the CNRS/IN2P3 Computing Center (Lyon - France) for providing computing and data-processing resources needed for this work.
The authors acknowledge the use of the Canadian Advanced Network for Astronomy Research (CANFAR) Science Platform operated by the Canadian Astronomy Data Centre (CADC) and the Digital Research Alliance of Canada (DRAC), with support from the National Research Council of Canada (NRC), the Canadian Space Agency (CSA), CANARIE, and the Canada Foundation for Innovation (CFI).
LWKG thanks the University of Edinburgh School of Physics and Astronomy for a postdoctoral Fellowship.
CD and MK acknowledge support from the Agence Nationale de la Recherche (ANR-22CE31-0014-01) TOSCA project.
FHP acknowledges support from CNES.
H. Hildebrandt is supported by a DFG Heisenberg grant (Hi 1495/5-1), the DFG Collaborative Research Center SFB1491, an ERC Consolidator Grant (No. 770935), and the DLR project 50QE2305.
MJH acknowledges support from NSERC through a Discovery Grant.
LVW acknowledges support from NSERC through a Discovery Grant. 
AHW is supported by the Deutsches Zentrum für Luft- und Raumfahrt (DLR), under project 50QE2305, made possible by the Bundesministerium für Wirtschaft und Klimaschutz, and acknowledges funding from the German Science Foundation DFG, via the Collaborative Research Center SFB1491 "Cosmic Interacting Matters - From Source to Signal".
We would like to thank our external blinding coordinator, Koen Kuijken.
\end{acknowledgements}

\bibliographystyle{aa} 
\bibliography{2D_harmony_paper} 

\begin{appendix}

\section{Impact of the point spread function on the cosmic shear power spectrum}
\label{app:PSF_impact}
\begin{figure}
    \centering
    \includegraphics[width=1\linewidth]{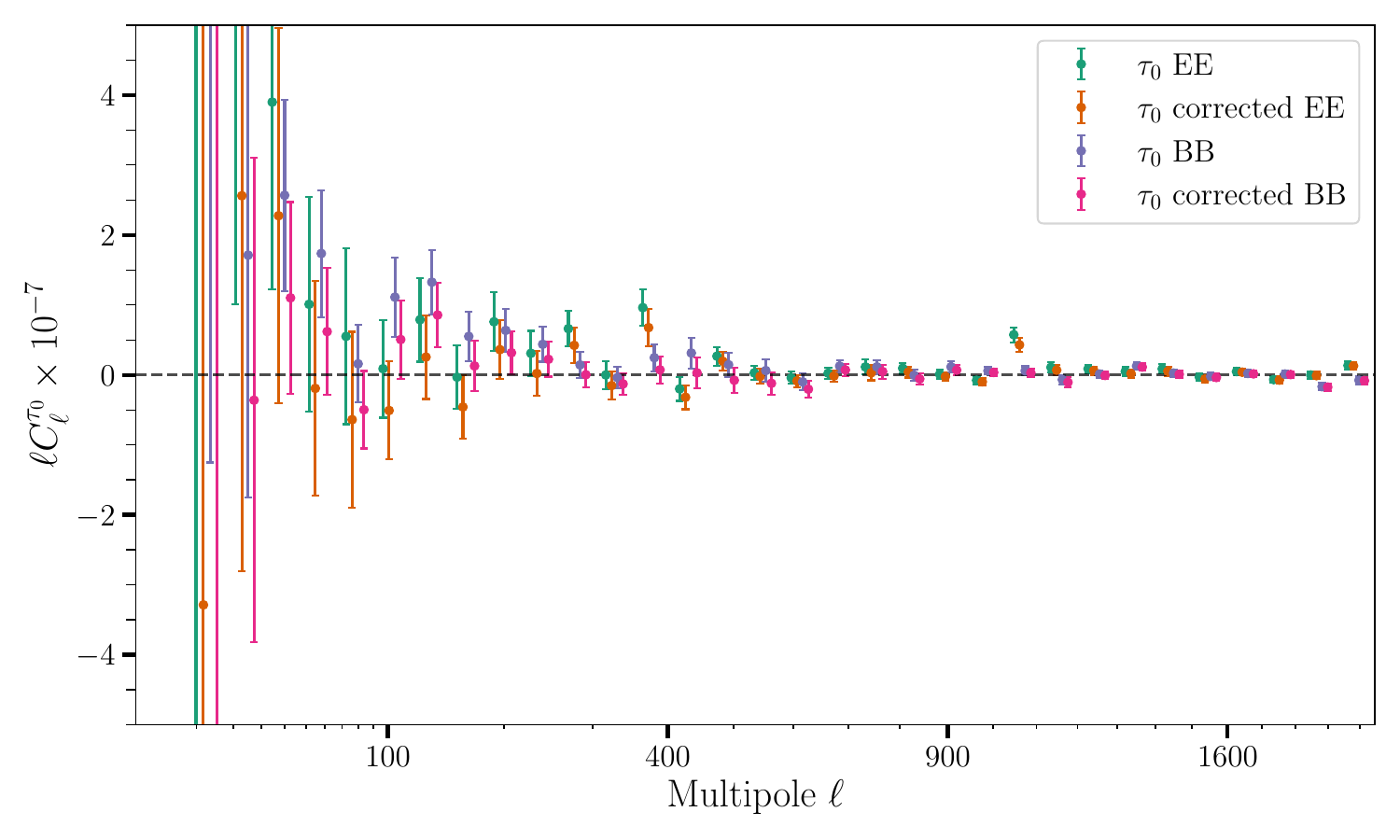}
    \caption{Galaxy-PSF power spectrum introduced in Eq.~\eqref{eq:tau_0_cl}. The $EE$ and $BB$ parts of the power spectrum are shown for raw ellipticities in the shape catalogue and for the object-wise leakage-corrected ellipticities using the methodology in \cite{liKiDSLegacyCalibrationUnifying2023} and described in Hervas-Peters et al., in prep. The signal mostly appears on large scales, which is consistent with configuration space leakage measurements. The leakage correction reduces the amplitude of the galaxy-PSF correlation in both $E$ and $B$ modes. Data points are slightly offset for visualisation purposes.}
    \label{fig:tau_0_cell}
\end{figure}

\begin{figure}
    \centering
    \includegraphics[width=1\linewidth]{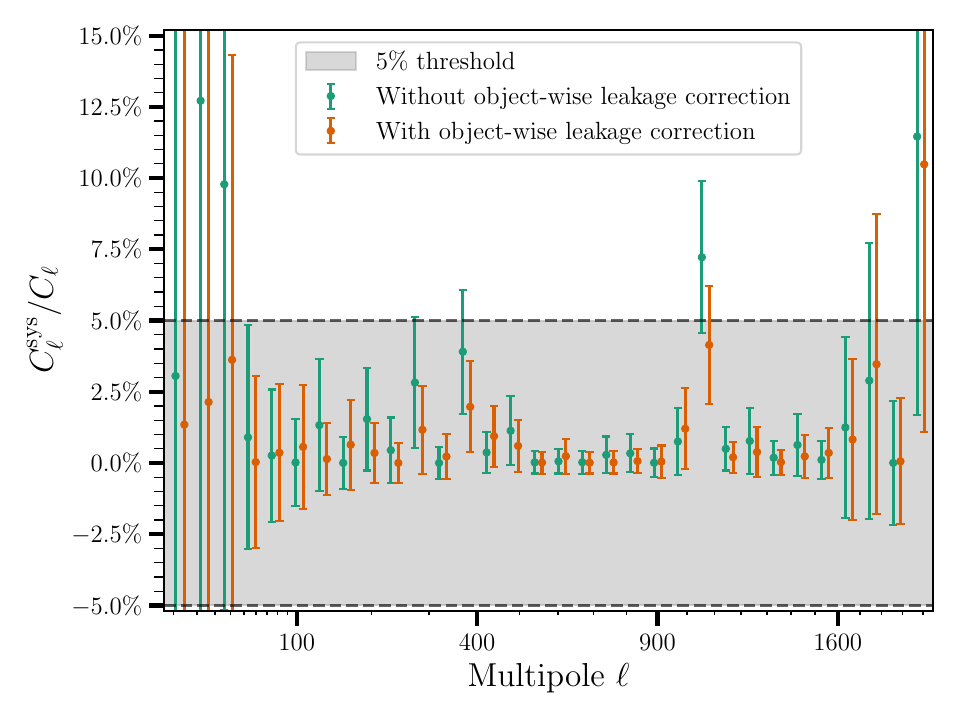}
    \caption{Ratio between the estimated additive PSF systematics contribution, $C_\ell^\mathrm{sys}$ (see Eq.~\eqref{eq:cl_sys}), and the cosmological signal $C_\ell^{EE}$. The ratio is shown for raw ellipticities in the shape catalogue and for the object-wise leakage-corrected ellipticities. The leakage correction reduces the amplitude of the systematic contribution to below 5\% on almost all scales and below 2.5\% for most multipoles. Data points for the leakage-corrected case are slightly offset for visualisation purposes.}
    \label{fig:ratio_cell_cell_sys}
\end{figure}

The primary source of PSF systematics in configuration space is associated with PSF leakage, which corresponds to a non-zero correlation between galaxy and PSF ellipticities. We assume that the observed ellipticities, defined in Sect.~\ref{sec:pseudo_cl}, follow
\begin{align}
    \bs{e}^\mathrm{obs} = \bs{\epsilon}^\mathrm{s} + \bs{\gamma} + \alpha \bs{e}^\mathrm{PSF},
\end{align}
where $\bs{\epsilon}^{\rm s}$ is the intrinsic ellipticity of the galaxy, $\bs{e}^\mathrm{PSF}$ is the ellipticity of the PSF at the galaxy position and $\alpha$ quantifies the amplitude of the PSF leakage and is modelled as a scalar quantity. We can then compute the equivalent of $\rho_0(\vartheta)$ and $\tau_0(\vartheta)$ \citep{jarvisDarkEnergySurvey2021, gattiDarkEnergySurvey2021},
\begin{align}
    C_\ell^{\rho_0} &= \langle \bs{e}^\mathrm{PSF} \bs{e}^\mathrm{PSF} \rangle_\ell,\\
    \label{eq:tau_0_cl}
    C_\ell^{\tau_0} &= \langle \bs{e}^\mathrm{obs} \bs{e}^\mathrm{PSF} \rangle_\ell.
\end{align}
Following the configuration-space methodology in \cite{kilbingerUNIONSWeakLensing2026}, we can estimate the amplitude of the contribution of PSF systematics to the observed cosmic shear power spectrum,
\begin{align}
    \label{eq:cl_sys}
    C_\ell^\mathrm{sys} \equiv \frac{({C_\ell^{\tau_0}})^2}{C_\ell^{\rho_0}}.
\end{align}
Figure~\ref{fig:tau_0_cell} shows the galaxy-PSF power spectrum $C_\ell^{\tau_0}$. We see that most of the signal is located on large scales (small multipoles), which corresponds to the behaviour of the leakage in configuration space \citep[see][]{kilbingerUNIONSWeakLensing2026,gohUNIONSWeakLensing2026}. The $EE$ and $BB$ parts of the power spectrum are represented for raw ellipticities in the shape catalogue and for the object-wise leakage-corrected ellipticities using the methodology in \cite{liKiDSLegacyCalibrationUnifying2023}. We can note two things. First, the object-wise correction tends to reduce the amplitude of $C_\ell^{\tau_0}$ in both $E$ and $B$ modes, which shows the positive effect of the leakage calibration. Second, we see that the galaxy-PSF signal, concentrated in the `$+$' component of the correlation function in configuration space, splits into $E$ and $B$ modes with comparable amplitude. This means that PSF systematic effects generate some $B$-mode signal, but at the same time, the amplitude of PSF systematics present in the fitted $EE$ signal is reduced. This can be seen from Fig.~\ref{fig:ratio_cell_cell_sys}, showing the ratio between the estimated additive PSF systematics contribution, $C_\ell^\mathrm{sys}$ and the cosmological signal. We observe that after applying the leakage correction, the amplitude of the systematic contribution remains below 5\% on almost all scales and below 2.5\% for most of the multipoles, while the significance of this additive bias exceeded 10\% on large scales \citep[see][]{gohUNIONSWeakLensing2026}. In addition, we can note two peaks in the $EE$ part of $C_\ell^{\tau_0}$ (see Figure~\ref{fig:tau_0_cell}) and equivalently in $C_\ell^\mathrm{sys}$ at $\ell=400$ and $\ell=1000$. These correspond to peaks in the Bessel function of the Hankel transform at $\theta \sim 27'$ and $\theta \sim 10'$, respectively. We highlight that this excess at $\ell=1000$ corresponds to the excess observed in the two-point correlation function $\xi_+$ at $\theta=10'$ in \cite{gohUNIONSWeakLensing2026}. This peak is, however, not seen in the configuration-space $\tau_0$ correlation function \citep{kilbingerUNIONSWeakLensing2026}.  Uncertainties for $C_\ell^{\tau_0}$ are estimated from the iNKA estimator and are used to derive error bars of the additive systematic contribution, $C_\ell^\mathrm{sys}$. For the latter, the size of the error bars is underestimated because the estimation of $C_\ell^{\rho_0}$ covariance matrix is too unstable to be used. The error bars shown in Fig.~\ref{fig:ratio_cell_cell_sys} are then only informative, to give the reader an idea of the order of magnitude of the uncertainty.

From the above analysis, we consider that the amplitude of the PSF systematics contribution, $C_\ell^\mathrm{sys}$, is small enough to be neglected in modelling. In Sect.~\ref{sec:robustness_modelling_psf}, we perform verification inference runs to validate that the PSF systematics do not have a substantial impact on the inference.

\section{Density variation and covariance estimation}
\label{app:density_variation_covariance}

In Sect.~\ref{sec:cov_modelling}, we showed that the iNKA covariance used in this work overestimates by around $20\%$ the uncertainty compared to what is obtained from \texttt{OneCovariance} and \texttt{GLASS} simulations. Among the differences between the three methods, iNKA differs in its treatment of the noise. An underlying assumption in \texttt{OneCovariance} and \texttt{GLASS} mocks is that the noise is stationary. This is not true if the object density varies strongly across the sky. By randomly rotating galaxies in the UNIONS lensing sample and projecting this on the sphere, we account for such variations in the noise bias term.

\begin{figure*}
    \centering
    \includegraphics[width=1\linewidth]{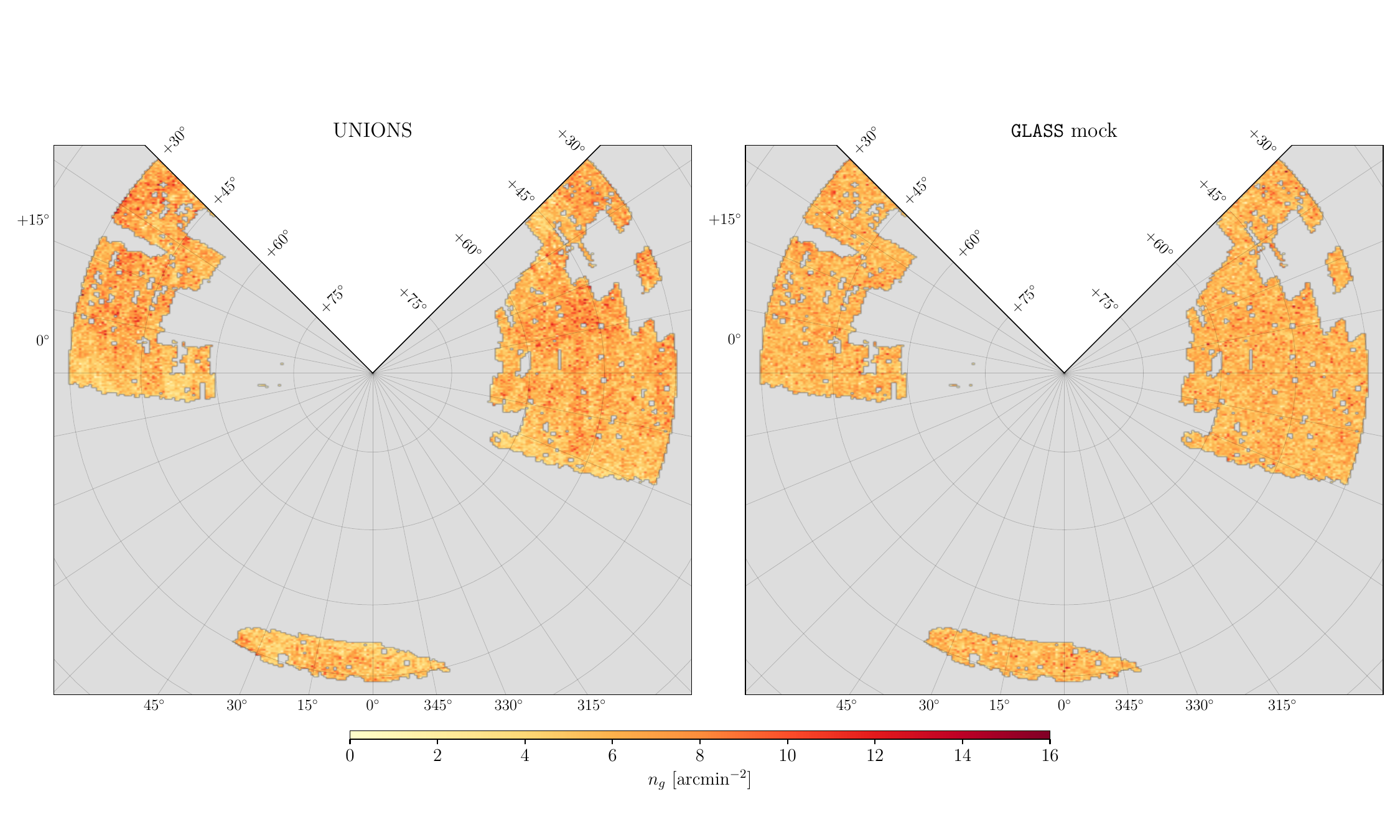}
    \caption{Comparison of the density map of UNIONS data (left) and a \texttt{GLASS} mock (right). The \texttt{GLASS} mock is statistically homogeneous by construction and allows us to assess the inhomogeneity of UNIONS data by eye. UNIONS data shows a reduced density near the Galactic plane, likely due to crowding and dust extinction.}
    \label{fig:density_map_comparison}
\end{figure*}

We test the impact of density variation on the covariance estimation by running iNKA on a \texttt{GLASS} mock where the galaxy sampling is homogeneous and does not suffer from depth variations. Figure~\ref{fig:density_map_comparison} shows a comparison of the density map of UNIONS data and a \texttt{GLASS} mock. We observe that the density of galaxies near the Galactic plane is lower in UNIONS data. This effect may be due to crowding from the high stellar density in this region of the sky or to dust extinction.

\begin{figure}
    \centering
    \includegraphics[width=1\linewidth]{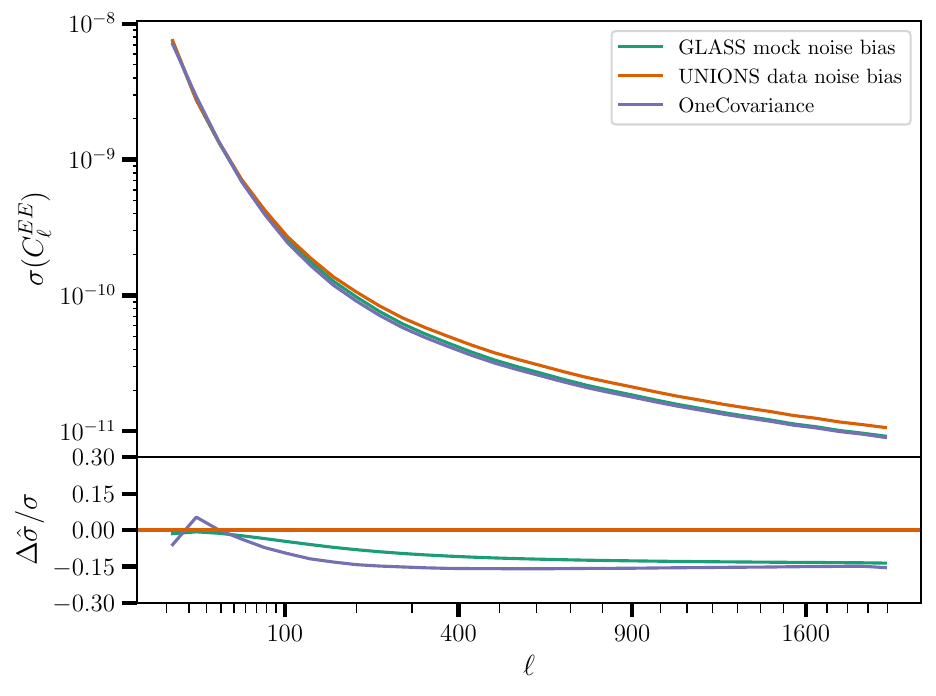}
    \caption{Comparison of the error bars obtained from iNKA covariance estimation using the noise bias of UNIONS data (orange line) and a \texttt{GLASS} mock with homogeneous density (green line). Using a homogeneous density reduces the error bars by about $20\%$, demonstrating the impact of density variations on covariance estimation. This accounts for the observed mismatch between \texttt{OneCovariance} and iNKA.}
    \label{fig:density_variation_effect}
\end{figure}

Figure~\ref{fig:density_variation_effect} shows the uncertainty obtained from iNKA with the noise bias estimated from UNIONS data and from a \texttt{GLASS} mock with homogeneous density. The about $20\%$ discrepancy between methods reflects spatially varying shape noise, which iNKA captures and \texttt{OneCovariance} does not; as a result we adopt iNKA as the fiducial covariance.

\section{Validation of the $B$-mode covariance}
\label{app:B_mode_cov_validation}

\begin{figure}
    \centering
    \includegraphics[width=1\linewidth]{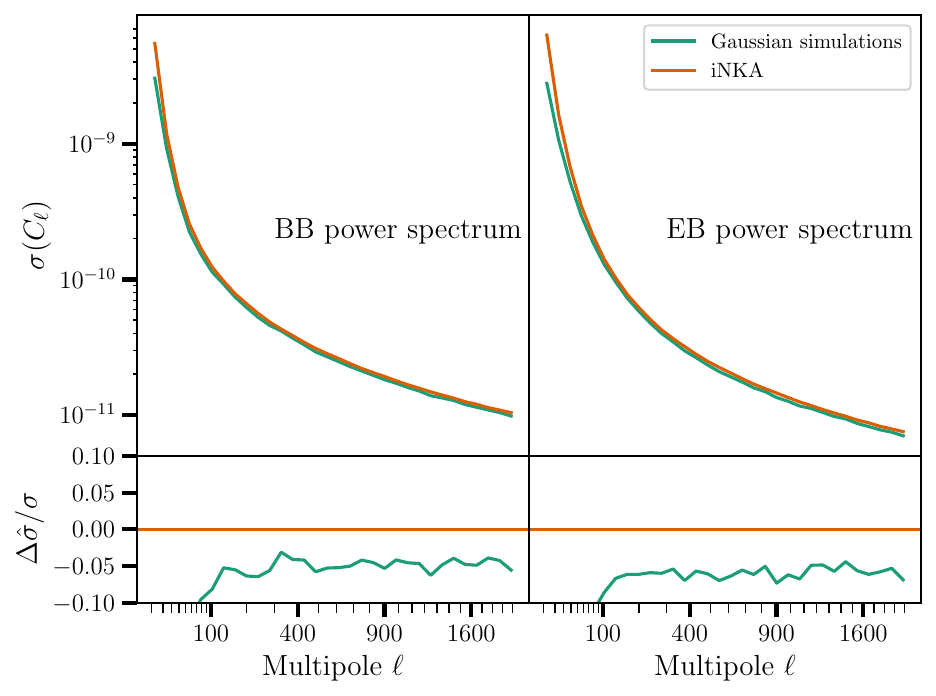}
    \caption{Comparison of the standard deviation of the $BB$ power spectrum obtained from $10 \,000$ Gaussian simulations (green line) and the iNKA covariance estimation (orange line). Both agree at the $5\%$ level on intermediate and small scales. However, on the largest scales, iNKA overestimates the uncertainty by more than $10\%$. }
    \label{fig:B_mode_cov_validation}
\end{figure}

In Sect.~\ref{sec:B_modes}, we measure and assess the significance of the $B$ modes present in our cosmic shear power spectrum measurement with UNIONS data. We estimated the covariance between the data points using the Gaussian covariance obtained with iNKA, like for the $EE$ part of the power spectrum. We validate this choice by measuring the $BB$ component of the power spectrum on $10 \, 000$ Gaussian simulations. A Gaussian signal is sampled from a fiducial power spectrum on top of which noise is added by randomly rotating galaxies from the UNIONS weak lensing sample, thus preserving the spatial variation of the shape noise in the data. The $BB$ power spectrum is then measured using the same pseudo-$C_\ell$ approach as for the data. Figure~\ref{fig:B_mode_cov_validation} shows a comparison of the standard deviation obtained from these $10 \, 000$ Gaussian simulations and the iNKA covariance estimation. The two approaches agree well, validating our choice to use iNKA for the $B$-mode covariance.

The agreement is at the $5\%$ level on scales larger than $\ell = 100$, and some discrepancy appears on the largest scales where iNKA overestimates the error bars. The discrepancy is due to an imperfect treatment of the $E/B$ mixing caused by the sky mask in the NKA estimator. Here we use the covariance computed from Gaussian simulations to assess the $EB$ and $BB$ significance reported in Sect.~\ref{sec:B_modes} and \cite{daleyUNIONSWeakLensing2026}. Using all scales, we find that the $EB$ power spectrum is consistent with the null hypothesis with a PTE of $0.82$. For the $BB$ power spectrum, we find a PTE of $0.013$ below our threshold of $p_\mathrm{thr}=0.05$ to claim a significant detection of $B$ modes. Applying the scale cuts described in Sect.~\ref{sec:scale_cut}, the PTE for the $BB$ power spectrum increases to $0.13$, showing that the $B$ modes are not significant when removing the largest and smallest scales and further motivating our choice of scale cuts. The $EB$ power spectrum PTE after the cuts amounts to $0.40$. Our fiducial inference setup is thus robust to the choice of covariance when assessing the significance of $B$ modes.

\section{Scale cut to mitigate sensitivity to baryonic feedback}
\label{app:scale_cut}

Section~\ref{sec:cs_power_spectrum} highlighted the difficulty of modelling the non-linear part of the power spectrum, as well as the effect of baryonic feedback. To make the analysis robust to this uncertainty, we remove the scales where our theoretical model fails to accurately capture the physical effects. Our method derives an approximate cut in multipoles from a small-scale cut of 3D Fourier modes, motivated by theoretical considerations. In practice, we assume that the theoretical model is valid up to a specific wavenumber $k_\mathrm{max}$ and we discard the multipoles $\ell$ that receive significant contributions from smaller scales, $k > k_\mathrm{max}$.

\begin{figure}
    \centering
    \includegraphics[width=1\linewidth]{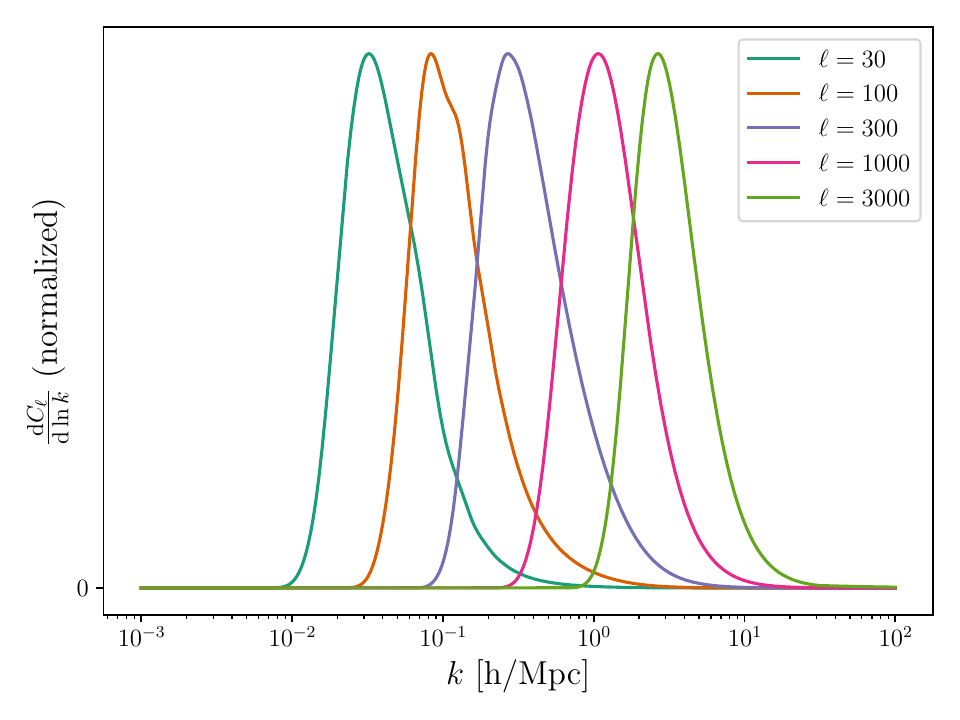}
    \caption{ Integrand of the cosmic shear power spectrum with respect to $k$ in $h \, {\rm Mpc}^{-1}$. For higher multipoles, higher wavenumbers contribute to the integrand, so smaller scales are captured by the multipole in the angular power spectrum. Section~\ref{sec:scale_cut} develops how these curves can be used to derive a scale cut in harmonic space given a maximum wavenumber $k_\mathrm{max}$.}
    \label{fig:Cl_integrand_vs_k}
\end{figure}

Following \cite{douxConsistencyCosmicShear2021}, we rewrite Eq.~\eqref{eq:cs_power_spectrum} as an integral over $k$ modes. We then define the scale $k_{>\alpha}(\ell)$ at which the integral for $C_\ell$ reaches a fraction $\alpha$ of its total value, such that
\begin{align}
\label{eq:Cl_integrand}
    \int_{-\infty}^{\ln k_{>\alpha}(\ell)} \drom \ln k \frac{\drom C_\ell}{\drom \ln k} = \alpha C_\ell. 
\end{align}
Figure~\ref{fig:Cl_integrand_vs_k} shows the integrand $\drom \ln C_\ell / \drom \ln k$ varying with the wavenumber $k$ in $h \, {\rm Mpc}^{-1}$. This shows that estimating $k_{>\alpha}(\ell)$ from Eq.~\eqref{eq:Cl_integrand} corresponds to finding the wavenumber for which the integral under the curves in Fig.~\ref{fig:Cl_integrand_vs_k} amounts to $\alpha$\% of the total value. For a given choice of $\alpha$ and $k_\mathrm{max}$, we then obtain the small-scale multipole cut by numerically solving for $\ell_\mathrm{max}$ such that $k_{>\alpha}(\ell_\mathrm{max}) = k_\mathrm{max}$. We set $\alpha=0.95$, i.e., wavenumbers $k$ larger than $k_{>\alpha}(\ell)$ contribute at most 5\% of the total signal, and consider $k_\mathrm{max} = 1, 3$ and $5$ $h \, {\rm Mpc}^{-1}$ cuts. This scale cut definition is model-dependent because the sensitivity to some $k$-modes depends on the shape of the 3D matter power spectrum. In Sect.~\ref{sec:robustness_modelling}, we explore the impact of varying the non-linear matter power spectrum model while preserving our scale cuts, including the sensitivity of baryonic feedback modelling in \texttt{HMCode2020} on our results. Solving for $\ell_\mathrm{max}$, and given our redshift distribution, we find that the scale cut associated with $k_\mathrm{max} = 3 h \, {\rm Mpc}^{-1}$ is $\ell_\mathrm{max} = 1800$.

\section{Additional results}
\label{app:additional_results}

\subsection{Full posterior}

\begin{figure*}
    \centering
    \includegraphics[width=1\linewidth]{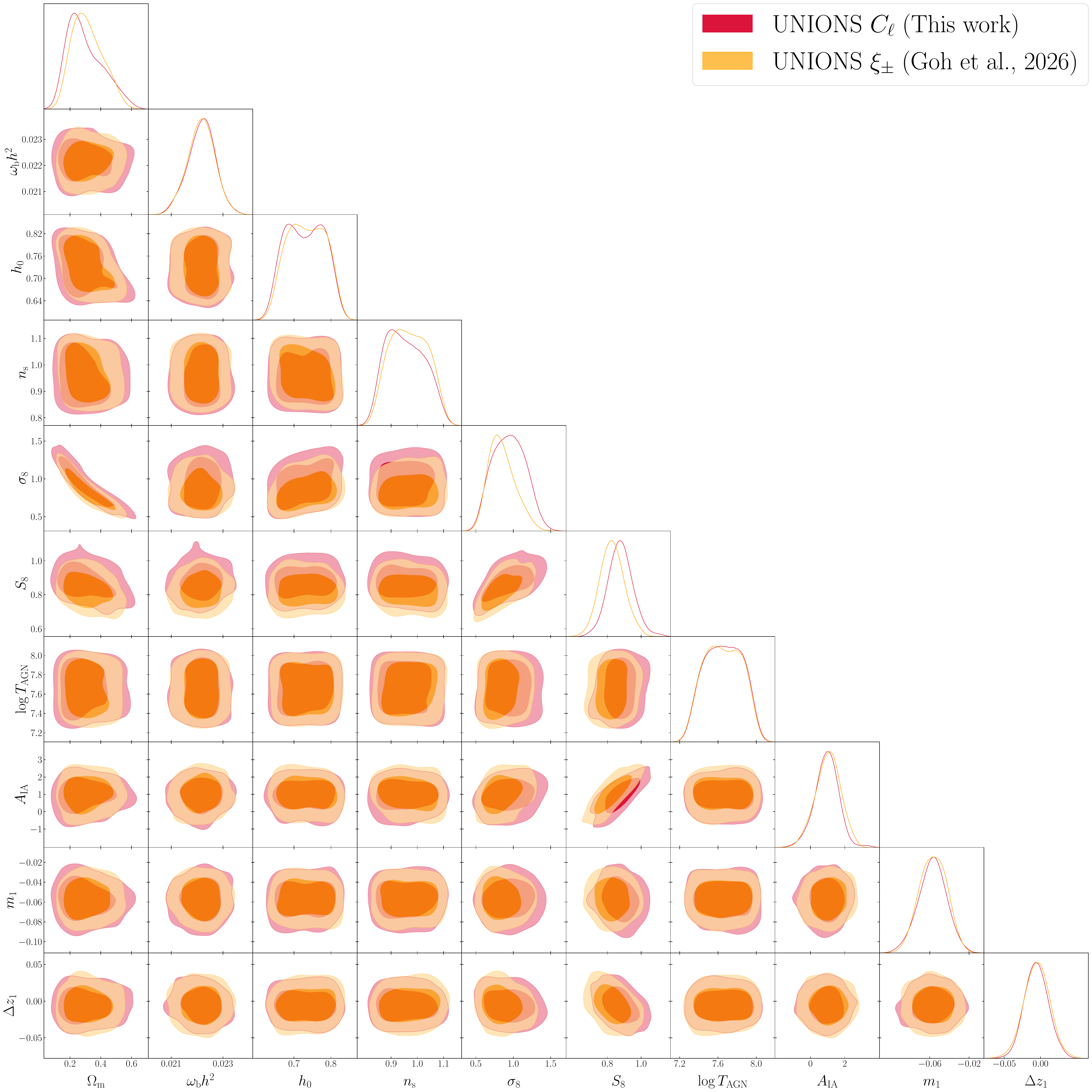}
    \caption{Full posterior obtained in harmonic space (red contour) and configuration space (orange contour). We highlight the strong degeneracy between the amplitude of the growth of structure, $S_8$, and of intrinsic alignment, $A_{\rm IA}$ due to the inability of our non-tomographic analysis to break degeneracies between the two effects.}
    \label{fig:full_posterior}
\end{figure*}

Figure~\ref{fig:full_posterior} shows the complete posterior obtained in harmonic and real space. Other parameters than $S_8$ and $\Omega_{\rm m}$ are not discussed in the main text but are summarised in Table~\ref{tab:metrics}, along with goodness-of-fit metrics. From this full posterior, we see that nuisance parameters are not constrained by our analysis. The redshift distribution shift, $\Delta z$, and the multiplicative shear bias, $m$, still follow their respective priors. The intrinsic alignment amplitude, $A_\mathrm{IA}$, exhibits a strong degeneracy with $S_8$ as described in Sect.~\ref{sec:intrinsic_alignment_prior}. Increasing the uncertainty in the amplitude of the intrinsic alignment prior would thus increase the uncertainty on $S_8$, hence the conservative choice made in Sect.~\ref{sec:intrinsic_alignment_prior}.

\subsection{Parameter point estimate}
\label{app:point_estimate}

The parameter point estimate reported in the main text is obtained using the maximum a posteriori (MAP). It is evaluated using a kernel density estimation (KDE) of the posterior distribution. The KDE interpolates the 1D marginalised posterior of each parameter except for $S_8$ and $\Omega_{\rm m}$, where the KDE is performed on the 2D marginalised posterior. This choice is motivated by degeneracies between $S_8$ and $\Omega_{\rm m}$ responsible for projection effects and the fact that using the 2D marginalised posterior gives a better agreement with the input cosmology in the validation test performed on the \texttt{GLASS} mocks compared to using the 1D marginalised posterior or the weighted average of the \texttt{Polychord} samples.

\subsection{Blinding}
\label{app:blinding}

To prevent confirmation bias, we made our analysis choices in a blinded setup using three different redshift distributions, without knowing which was correct. Blinding was performed by an external collaborator who applied a random shift to the redshift distribution, thus inducing an offset of the marginalised $S_8$ posterior distributions. The unknown modification creates a shift in $S_8$ (see Fig.~\ref{fig:whisker_plot}). Note that the three blinds span $S_8 \in [0.837,0.891]$ which is a smaller range than the data uncertainty $\sigma(S_8) \simeq 0.07$. The bulk of this manuscript was prepared prior to unblinding and underwent review by the broader UNIONS collaboration, including the external blinding coordinator. During post-processing of the inference chains, the primary authors compared the maximum-likelihood samples of the three blinds, which revealed a coherent preference for a single S8 value prior to the formal unblinding step. We also performed an updated run of inference chains after unblinding, moving the mean redshift distribution shift, $\Delta z$, from a previously wrong value of $-0.030$ to a corrected value of $-0.003$. This modification was implemented because of a correction of an error in the pipeline used to estimate $\Delta z$. Our inference results shifted towards lower $S_8$ values by about $0.03$. That being said, no other inputs to the inference were modified after unblinding; only the corrected $\Delta z$ prior was used, with new chains run accordingly.

\renewcommand{\arraystretch}{1.3} 

\begin{table*}
\begin{tabular}{l|c|c|c|c|c|c|c}
\hline
\hline
Experiment name & $S_8$ & $\Omega_m$ & $\sigma_8$ & $A_\mathrm{IA}$ & $\log T_\mathrm{AGN}$ & $\chi^2$ & $N_{\rm data}$ \\ 
\hline
UNIONS $C_\ell$ (This work) &  $0.891^{+0.057}_{-0.084}$ &  $0.225^{+0.153}_{-0.077}$ &  $0.960^{+0.223}_{-0.257}$ &  $1.011^{+0.655}_{-0.694}$ &  $7.627^{+0.263}_{-0.187}$ & 8.06 & 17 \\
UNIONS $\xi_\pm(\vartheta)$, (Goh et al., 2026) &  $0.831^{+0.067}_{-0.078}$ &  $0.265^{+0.130}_{-0.075}$ &  $0.782^{+0.206}_{-0.144}$ &  $1.128^{+0.688}_{-0.798}$ &  $7.557^{+0.322}_{-0.127}$ & 9.50 & 14 \\
$k_{\rm max}=5h\,\mathrm{Mpc}^{-1},\ \ell_{\rm max}=2048$ &  $0.889^{+0.060}_{-0.075}$ &  $0.218^{+0.150}_{-0.071}$ &  $0.963^{+0.217}_{-0.249}$ &  $1.057^{+0.720}_{-0.700}$ &  $7.835^{+0.111}_{-0.302}$ & 10.02 & 21 \\
$k_{\rm max}=3h\,\mathrm{Mpc}^{-1},\ \ell_{\rm max}=1800$ &  $0.885^{+0.072}_{-0.078}$ &  $0.208^{+0.162}_{-0.060}$ &  $1.004^{+0.184}_{-0.257}$ &  $1.041^{+0.757}_{-0.799}$ &  $7.802^{+0.093}_{-0.340}$ & 9.89 & 19 \\
$k_{\rm max}=1h\,\mathrm{Mpc}^{-1},\ \ell_{\rm max}=500$ &  $0.872^{+0.058}_{-0.092}$ &  $0.257^{+0.187}_{-0.073}$ &  $0.758^{+0.272}_{-0.158}$ &  $1.013^{+0.688}_{-0.747}$ &  $7.485^{+0.382}_{-0.079}$ & 1.71 & 4 \\
Include large scales, $\ell_{\rm max}=1600$ &  $0.890^{+0.072}_{-0.076}$ &  $0.235^{+0.135}_{-0.071}$ &  $0.894^{+0.247}_{-0.166}$ &  $1.181^{+0.666}_{-0.620}$ &  $7.536^{+0.297}_{-0.117}$ & 14.64 & 28 \\
Small scales only &  $0.877^{+0.061}_{-0.100}$ &  $0.237^{+0.184}_{-0.066}$ &  $0.746^{+0.334}_{-0.122}$ &  $0.967^{+0.718}_{-0.710}$ &  $7.812^{+0.084}_{-0.362}$ & 5.64 & 9 \\
Large scales only &  $0.876^{+0.064}_{-0.081}$ &  $0.233^{+0.215}_{-0.040}$ &  $0.765^{+0.274}_{-0.138}$ &  $0.862^{+0.792}_{-0.614}$ &  $7.785^{+0.102}_{-0.383}$ & 5.43 & 8 \\
\texttt{Halofit} &  $0.851^{+0.067}_{-0.067}$ &  $0.214^{+0.167}_{-0.066}$ &  $0.954^{+0.180}_{-0.261}$ &  $1.043^{+0.620}_{-0.774}$ &  N/A & 9.40 & 17 \\
\texttt{HMCode} no baryons &  $0.871^{+0.066}_{-0.082}$ &  $0.209^{+0.182}_{-0.061}$ &  $0.878^{+0.266}_{-0.204}$ &  $0.990^{+0.698}_{-0.706}$ &  N/A & 8.52 & 17 \\
\texttt{OneCovariance} only &  $0.886^{+0.060}_{-0.071}$ &  $0.206^{+0.144}_{-0.065}$ &  $0.961^{+0.245}_{-0.224}$ &  $0.998^{+0.697}_{-0.671}$ &  $7.808^{+0.112}_{-0.318}$ & 14.33 & 17 \\
No leakage correction &  $0.893^{+0.052}_{-0.098}$ &  $0.221^{+0.165}_{-0.071}$ &  $0.814^{+0.354}_{-0.131}$ &  $1.027^{+0.622}_{-0.826}$ &  $7.821^{+0.084}_{-0.353}$ & 8.53 & 17 \\
UNIONS $C_\ell$, Blind A &  $0.862^{+0.053}_{-0.083}$ &  $0.218^{+0.156}_{-0.071}$ &  $0.995^{+0.145}_{-0.311}$ &  $0.944^{+0.738}_{-0.709}$ &  $7.763^{+0.125}_{-0.351}$ & 7.68 & 17 \\
UNIONS $C_\ell$, Blind C &  $0.837^{+0.046}_{-0.071}$ &  $0.227^{+0.145}_{-0.080}$ &  $0.900^{+0.194}_{-0.240}$ &  $0.881^{+0.834}_{-0.606}$ &  $7.779^{+0.125}_{-0.307}$ & 7.46 & 17 \\
\hline
\end{tabular}
\caption{Summary of the 1D marginal constraints on $S_8$, $\Omega_{\rm m}$, $\sigma_8$, $A_\mathrm{IA}$ and $T_\mathrm{AGN}$ for our fiducial analysis and variations developed in Sect.~\ref{sec:robustness_modelling}. We also report the $\chi^2$ of the best-fit model and the number of data points used in the fit.}
\label{tab:metrics}
\end{table*}

\end{appendix}

\end{document}